# Improvements in Antineutrino Detector Response by Including Fission Product Isomeric Transitions and Corrections using New Data

Wei Eng Ang[a], Sanghun Lee[a], Shikha Prasad[a]*,

[a]*Texas A&M University, Department of Nuclear Engineering, College Station, Texas*

*E-mail: shikhap@tamu.edu

CEvNS detectors could provide new opportunities in nuclear physics applications and applications in nuclear engineering if they can improve existing parameters such as (anti)neutrino detector size, portability, detection times (intrinsic efficiency), their sensitivity to a large range of reactor antineutrino energies (threshold of detection), and resources required for operation. Thus, modelling the antineutrino spectrum is a crucial step to study the reactor antineutrino spectra and the CEvNS detector response. The first objective of this paper is to study the importance of fission product libraries in the construction of antineutrino spectrum using the summation method and with various other corrections, including isomeric transitions in fission products, finite size, weak magnetism, and radiative corrections. We have used ENDF/B-VIII.0 and JEFF-3.3 nuclear libraries as our base data to model the antineutrino spectrum. We have also included the total absorption gamma spectroscopy (TAGS) data, which is free from the pandemonium effect, when such data is available. Our analysis includes the newest TAGS data sets from J. Gombass *et al.* (2021) with additions made after M. Estienne *et al.* (2019) reactor antineutrino spectra study involving TAGS data. The isomeric transitions correction has the highest impact on the antineutrino energy spectra increasing the values 29% to 37% on an average in the energy range of 0.5 MeV to 2 MeV. This antineutrino spectra correction also shows an increase of 4.71% to 7.13% in the range of 0 to 2 MeV, with improving isomeric transitions using the TAGS data from published literature. Next, antineutrino spectra including the isomeric-transition-correction using the Gross Theory causes reduction by 11.56 % to 69.46 % for all four fissionable isotopes in the range of 6 to 8 MeV. The finite size correction, radiative correction, and weak magnetism corrections cause no more than 3.27% difference between the corrected and uncorrected spectra. We studied the impact of various corrections to the antineutrino spectra and quantified the improvements made in the antineutrino spectrum calculation due to these changes. However, we have not included forbidden decays to simplify the calculations. The second objective of this paper is to calculate pulse height distributions of Ge- and Si-based CEvNS sensors assuming a 20-eV nuclear recoil threshold. Towards this objective, the probability that an antineutrino incident with a given energy, $E_\nu$ will cause a certain amount of nuclear recoil energy, $T$, in the detector is calculated to model pulse height distributions. These probability distributions are shown for several different incident antineutrino energies. In our study, we have assumed a 100 kg CEvNS detector placed 10 m away from the 1 MW TRIGA reactor with a 20-eV nuclear recoil energy threshold. Our results show that the detector response with corrected spectra for a CEνNS based natural Ge detector is 44.25 events/day and a natural Si detector is 7.99 events/day. As mentioned above, the biggest impact on the detector response is due to the isomeric transitions correction; a 37 % difference between the corrected and uncorrected detector response is observed.

Keywords: reactor antineutrino measurement; coherent elastic neutrino nucleus scatter; CEνNS; semiconductor detectors, pulse height distribution

**Table of Contents**

## 1. Introduction

The first observation of coherent-elastic-neutrino-nucleus-scattering (CE$\nu$NS) in 2017 has offered many possibilities to further nuclear physics frontiers and applications such as alternative nuclear safeguards possibilities using reactor antineutrinos [1] [2]. This is because the intrinsic cross-section of CE$\nu$NS is significantly higher than the current state-of-practice inverse beta decay (IBD), which utilizes the reaction as shown below [3]:

$$\bar{\nu}_e + p \rightarrow e^+ + n \tag{1}$$

where an antineutrino interacts with a proton to produce a positron and a neutron. Since the CE$\nu$NS has a higher probability to interact with antineutrinos, intrinsic detection efficiency can be improved and a smaller detector size can be achieved [4].

In a previous study, we demonstrated that a CE$\nu$NS based semiconductor detector has greater intrinsic detection efficiency compared to an IBD based detector [4]. However, CE$\nu$NS events lead to small energy depositions which are difficult to detect. New technological breakthroughs such as the Z-sensitive Ionization and Phonon (ZIP) detector and the Super Cryogenic Dark Matter Search (SuperCDMS) have enabled detecting these signatures [5]. Therefore, we explore the application of CE$\nu$NS based semiconductor detectors for a reactor antineutrino measurement, as such a measurement remains to be accomplished [2].

Modeling and understanding a source detector scenario requires accurate knowledge of both the source and the detector. In antineutrino detection, the source term determination is subject to large uncertainties emanating from nuclear data [6]. However, in nonproliferation analysis, a precise reactor antineutrino energy spectrum calculation is highly desirable to obtain an accurate detector response. Thus, it is imperative to understand the sources of uncertainties in the calculation of the antineutrino spectrum using the summation method and the nuclear data libraries. Evaluating these uncertainties is the objective of our study [7].

## 2. Literature review and previous works

### 2.1 Summation Method and Beta Spectrum Conversion Method

Currently, there are two ways to model the reactor antineutrino spectrum: the "ab-initio" summation method and the beta spectrum conversion method [8]. In the beta spectrum conversion method, an experimentally determined beta spectrum from the reactor core is used to determine the antineutrino spectrum. This is usually done by using the beta spectra that have been measured at the Institut Laue-Langevin (ILL) in Grenoble, France. Since individual $\beta$-branches information is not available from the measured integral beta spectrum, Schreckenbach *et al.* developed a conversion method to obtain 30 virtual $\beta$ branches by dividing the beta spectrum [9]. The $\beta$-branches are referred to as beta intensities of isomeric states as the beta spectrum is continuous, so the Q-values are different for each isomeric state. Then, the antineutrino spectrum is modelled by summing all 30 virtual antineutrino spectra by replacing $E_e = E_0 - E_\nu$, where $E_e$ is total electron energy, $E_0$ is the endpoint energy, and $E_\nu$ is antineutrino energy. The conversion method has been revisited by Mueller *et al.* (2011) [10] and Huber *et al.* (2011) [11] to predict antineutrino spectra for $^{235}$U, $^{239}$Pu and $^{241}$Pu. The beta spectrum of $^{238}$U was measured for the first time at Forschungsreaktor München II (FM II) in Garching bei München, Germany in 2013. Haag *et al.* has presented the $^{238}$U antineutrino spectrum using conversion method in 2013 [12]. The beta spectrum conversion method is considered a more precise measurement method to calculate the antineutrino spectrum compared to the summation method and will serve as our benchmark [13]. However, the conversion data is only available in the energy range of 2 to 8 MeV [10].

On the other hand, the summation method relies on the information from nuclear data libraries and models the antineutrino spectral from all the contributing fission products. The summation method can be used to calculate the low energy antineutrino spectra, but this method often overestimates the higher part energy of the antineutrino spectrum due to the pandemonium effect [14]. The summation method also suffers from large uncertainties in the cumulative yield, Q values, and beta intensities [13]. Reactor antineutrino spectral prediction using summation method has been performed by Vogel *et al.* (1985) [15] and Hayes. *et al.* (2016) [8]. Vogel *et al.* used the summation method to predict the antineutrino spectrum resulting from both thermal and fast fissions.

## 2.2 TAGS Data and Pandemonium Effect

When a fission product β decays with a large amount of available Q-value, the resulting daughter nuclide can have several different excitation energies or isomeric states available, resulting in high energy gamma-rays which take away from the kinetic energy of the beta-particle. However, it is well known that measuring high energy gamma-rays is a challenge since not only are these gamma-rays less frequent, but detection efficiency of gamma-rays at higher energies decreases exponentially as a function of energy. Therefore, measurements using high-resolution gamma-ray detectors often over-estimate the contribution of lower energy states and assign them higher beta intensity values, called the pandemonium effect [16]. Figure 1 shows the pandemonium effect and how it affects the beta intensity values. The beta intensity is the probability of a β decay to a given isomeric state with its own unique Q-value.

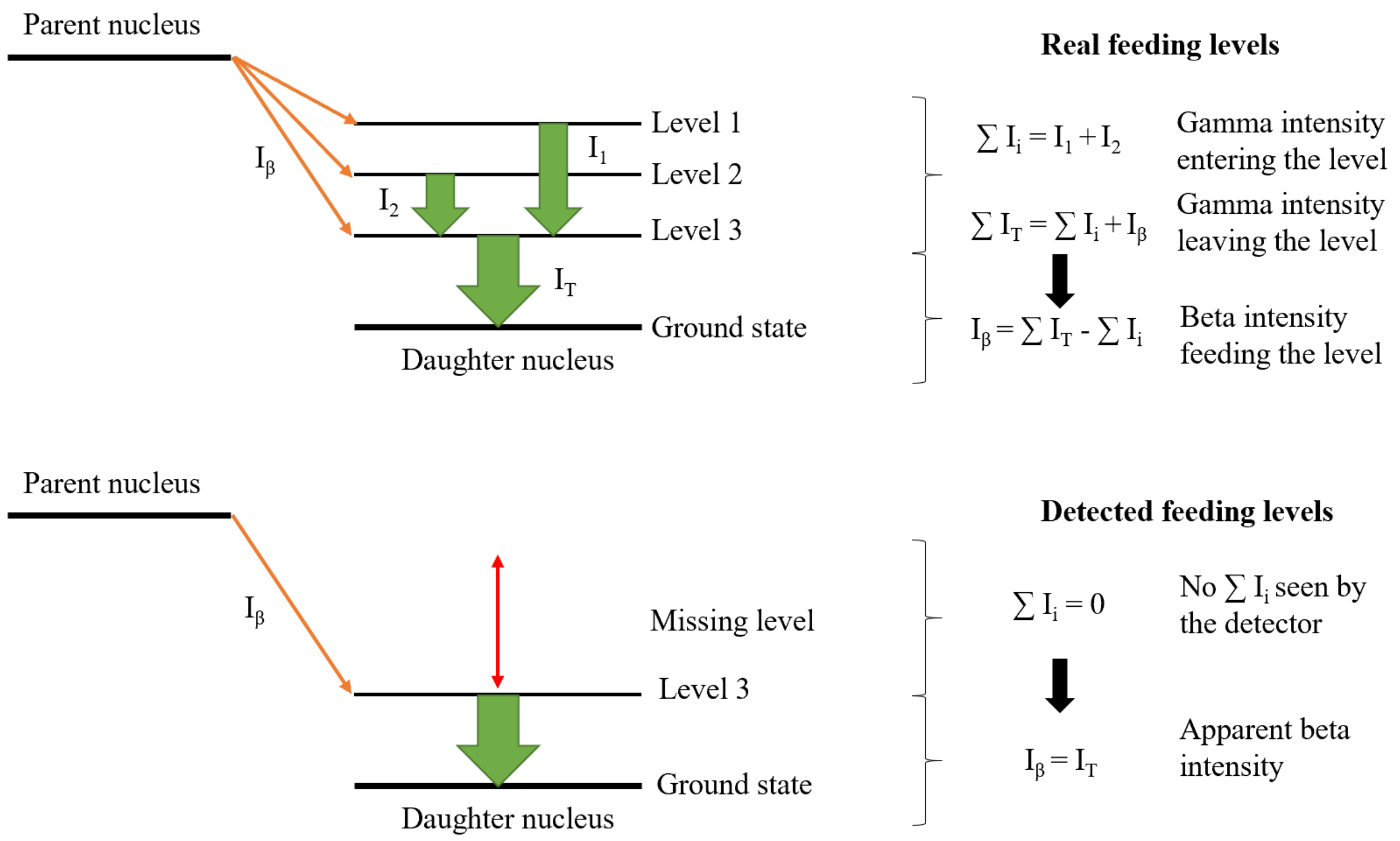

**Fig. 1**: *An illustration of the apparent pandemonium effect due to the inability to accurately measure gamma-rays and contributions from the higher isomeric states of a nucleus (Aguado and Esther, 2012).*

In this paper, we intend to simplify and model the reactor antineutrino spectrum using the summation method with latest nuclear libraries and the fission fragments data that have been measured in recent years using the total absorption gamma spectroscopy (TAGS) technique [17]. TAGS is a technique used to obtain beta intensity info from an isotope, which is free from the pandemonium effect [18]. Without using TAGS, the beta intensities can be determined incorrectly because the detection efficiency of high-purity-germanium (HPGe) detectors begins to decrease beyond 100 keV, which leads to missing information and incorrect calculation on the beta intensities. M. Fallot *et al.* (2017) and M. Estienne *et al.* (2019) have reconstructed the antineutrino spectrum using TAGS data [17] [19]. In this study, we have included the newest TAGS data with additions made after 2019's reactor antineutrino spectra study involving TAGS data [19].

Some of the beta intensity information, especially for those neutron-rich isotopes are not available in nuclear libraries such as ENDF/B-VIII.0 [20]. Thus, we use predicted continuous beta spectra from JENDL-2015 [21] and convert it into antineutrino spectra, which will be discussed in Section 3.2. Finally, we demonstrate the impact of including fission fragment transitions and various types of corrections in the summation method to the antineutrino detector response in Section 4. We also present pulse height distributions for germanium and silicon detectors in response to the reactor antineutrino spectrum from Texas A&M's 1 MW research reactor.

## 3. Methodology

In our study, we used the summation method to calculate the antineutrino spectra of $^{235}U$, $^{238}U$, $^{239}Pu$ and $^{241}Pu$ from various fissions fragments, using cumulative fission yields data from JEFF-3.3 [22] and decay data (such as Q values and beta intensities) from ENDF/B-VIII.0 [20]. Both fission yields and decay data were extracted and processed using both Python and Matlab programs. One of the challenges of using summation method is to obtain beta intensities of some neutron-rich isotopes because this information is not available in either ENDF/B-VIII.0 or ENSDF [23]. Several methods such as the quasiparticle random phase approximation (QRPA) model and statistical Hauser-Feshbach method [24] have been used to predict the beta intensities and delayed-neutron energy spectra. However, these methods often involve tedious calculations. Thus, for the sake of simplification, our study used continuous beta spectra from JENDL-2015 and converted them into antineutrino spectra. These continuous beta spectra are modeled using the Gross Theory, a theory based on the sum rule of beta strength to calculate beta decay rates and the delayed neutron in the nuclear mass region [25] [26]. Only those fission fragments with cumulative yield more than $10^{-6}$ [27] were selected in this study as the cumulative yield lower than that will not impact the spectra significantly. The priority order of data selection is used as below:

(i) Decay data from ENDF/B-VIII.0 [20]

(ii) The TAGS decay data sets [28] [29] [30] [31] [32] [33] [34]

(iii) Continuous beta spectra from JENDL -2015 [21]

The data from above are cross-checked with the ENSDF library to ensure the latest data is being used. If latest TAGS data are available, then they will be included in this analysis.

Lastly, one shall not ignore the contribution of the $^{238}$U breeding event as it also produces antineutrino during the neutron capture reaction. The $^{238}$U will first capture one neutron to produce $^{239}$U. Next, $^{239}$U will be beta decaying into $^{239}$Np and lastly to $^{239}$Pu with another beta decay reaction [35]. The contribution of $^{238}$U neutron capture to detector response and pulse height distribution will also be included in our study.

### 3.1 Antineutrino Spectrum Calculation and Corrections

An antineutrino spectrum of the fissionable isotopes ($^{235}$U, $^{238}$U, $^{239}$Pu and $^{241}$Pu) is the sum of all fission fragments that undergo $\beta$- decay with all their, $i$-th branches [15] [8]:

$$\frac{dN}{dE_v} = \sum_n Y_n(Z,A,t) \sum_i b_{n,i}(E_0^i) P_v(E_v, E_0^i, Z) \quad (2)$$

The beta intensities, $b_{n,i}$ are normalized to unity ($\sum_i b_{n,i} = 1$, unless the fission fragments have additional decay modes) and $P_v$ is the normalized antineutrino spectrum with endpoint energy, $E_0^i$. The $Y_n(Z,A,t)$ is number of $\beta$ decays of fission fragment Z and A at time t. After a reactor burns for sufficient time, $Y_n$ is approximately cumulative yield and independent of t. In this study, the fission fragments with cumulative yields greater than $10^{-6}$ were used in this study [27]. The antineutrino emission spectrum for a fission product can be written as [8]:

$$P_v\left(E_v, E_0^i, Z\right) = kp_e(E_0 - E_e)^2 F(Z, E_e) C(Z, E_e)[1 + \delta(Z, A, E_e)] \quad (3)$$

In our calculation, $k$ is the normalization constant, $p_e$ is electron momentum, $F(Z, E_v)$ is the Fermi function, Z is the atomic number, $E_v$ is the initial antineutrino energy incident on the detector, $E_e$ is the total electron energy, $m_0$is electron rest mass and $C(Z, E_e)$ is shape factor. For allowed decay, $C(Z, E_e)$ is equal to 1. The notation $E_0$ here is the endpoint energy and approximately equal to $Q + m_0 c^2$, where $Q$ is reaction Q-value. The function $\delta(Z, A, E_e)$ expresses the corrections that should be considered to the shape of the spectrum including finite size correction, radiative correction, and weak magnetism correction, which have been discussed in Huber *et al.* (2011) and Hayes *et al.* (2016) papers [8] [11]. To obtain antineutrino spectrum, Eq. (3) is required to substitute with $E_v = E_0 - E_e$.

The Fermi function, $F(Z, E_v)$ here is used to account for the Coulomb field on the shape of the beta spectrum. The Fermi Function is given by [8]:

$$F(Z, E_e) = 4(2p_e R)^{-2(1-\gamma)} \left|\frac{\Gamma(\gamma + iy)}{\Gamma(2\gamma + 1)}\right|^2 e^{\pi y} \quad (4)$$

where:

$\gamma = \sqrt{1 - (\alpha Z)^2}$ and $y = \propto ZE_e/p_e$. The $R$ is the nucleus radius and is given as $R = 1.2A^{1/3} fm$, $\alpha$ is the fine structure constant (~1/137), Z is the nuclear charge of daughter nucleus. The notation $i$ is an imaginary number and $\Gamma$ is the gamma function.

The finite size correction, $\delta_{FS}$, is used to correct for the electric charge spatial distribution and hypercharge distribution (from the point-charge treatment) of the nucleus [11]. The equation of a finite size correction is given by [36]:

$$\delta_{FS} = -\frac{8}{5}\frac{Z\alpha R E_e}{\hbar c}\left(1 + \frac{9}{28}\frac{m_e^2 c^4}{E_e^2}\right) \quad (5)$$

Where $\hbar$ is reduced Plank's Constant, $m_e$ the mass of electron, and $c$ is the speed of light.

Next, an approximation for weak magnetism correction is used in this study and defined as [8]:

$$\delta_{WM} = 0.5\ \%\ E_e\ MeV^{-1} \tag{6}$$

The weak magnetism correction is used to account the effect of the interaction of the outgoing electron with magnetic moment [8] [11]. Lastly, the radiative correction is used to account for the effect that arise from the interaction of the photon field and the electron-positron field [37]. The radiative correction can be defined as [8]:

$$\delta_{QED} = \frac{\alpha}{2\pi} h(\hat{E}, E_0) \tag{7}$$

The function $h(\hat{E}, E_0)$ has been computed by Sirlin (2011) is written as [38]:

$$\begin{aligned} h(\hat{E}, E_0) = 3\ln\left(\frac{m_p}{m_e}\right) + \frac{23}{4} - \frac{8}{\hat{\beta}} Li_2\left(\frac{2\hat{\beta}}{1+\hat{\beta}}\right) + 8\left(\frac{tanh^{-1}\hat{\beta}}{\hat{\beta}} - 1\right) ln\left(\frac{2\hat{E}\hat{\beta}}{m}\right) \\ + 4\frac{tanh^{-1}\hat{\beta}}{\hat{\beta}}\left[\frac{7+3\hat{\beta}^2}{8} - 2tanh^{-1}\hat{\beta}\right] \end{aligned} \tag{8}$$

where $\hat{E} = E_0 - E_v, \hat{p} = \left(\hat{E}^2 - m_e^2\right)^{1/2}$, and $\hat{\beta} = \hat{p}/\hat{E}$. The function $L_2(x)$ is the dilogarithm and defined as $L_2(x) = -\int_o^x (\frac{dt}{t}) \ln(1-t)$. Huber *et al.* has done a detail analysis on finite size, radiative and weak magnetism correction with the antineutrino spectra using conversion method [11].

### 3.2 Conversion of Continuous Beta Spectrum to Antineutrino Spectrum

For simplification, an empirical method is used to obtain antineutrino spectrum from continuous beta spectrum (JENDL-2015) by using equation below [12]:

$$N_v(E) = N_\beta(E - 511\ keV - 50\ keV) \cdot k(E) \tag{9}$$

The antineutrino spectra can be corrected by shifting the beta spectrum with the rest mass of the electron, and 50 keV to account the Coulomb attraction of nucleus and electron. The correction function $k(E)$ here is expected to be close to unity and in the order of 5 % [12].

### 3.3 CEvNS Cross-Section Calculation

The differentiate equation of CEvNS cross-section over the recoil energy is defined as [35]:

$$\frac{d\sigma(E_v)}{dT_R(E_v)} = \frac{G_F^2 M}{2\pi}[(q_v + q_A)^2 + (q_v - q_A)^2\left(1 - \frac{T_R}{E_v}\right)^2 - (q_v^2 - q_A^2)\frac{MT_R}{E_v^2}] \tag{10}$$

The Eq. 10 is integrated into Eq. 11 to obtain the CEvNS cross-section as a function of the incident antineutrino energy. The integrated CEvNS is shown as below:

$$\sigma(E_v) = \int_{T_R min}^{T_R max} A_1 T_R + A_2 T_R - \frac{A_2 T_R^2}{E_v} + \frac{A_2 T_R^3}{3{E_v}^2} - \frac{A_3 M T_R^2}{2{E_v}^2} \tag{11}$$

where $G_F$ denotes the Fermi constant, $M$ is the mass of targeted nucleus, $q_v$ and $q_A$ are vector and axial charges respectively, $T_R$ is recoil energy and $E_v$ is incident neutrino energy. The maximum recoil energy can be calculated by [35]:

$$T_R^{Max} = \frac{2E_v^2}{M + 2E_v} \tag{12}$$

$A_1$, $A_2$ and $A_3$ here are constants where:

$$A_1 = \frac{G_F^2 M}{2\pi}(q_v + q_A)^2 \tag{13}$$

$$A_2 = \frac{G_F^2 M}{2\pi}(q_v - q_A)^2 \tag{14}$$

$$A_3 = \frac{G_F^2 M}{2\pi}(q_v^2 - q_A^2) \tag{15}$$

The vector charges, $q_v$ is defined as:

$$q_v = g_v^p Z + g_v^n N \tag{16}$$

where $g_v^p$ and $g_v^n$ are vector proton and neutron weak neutral current couplings respectively where $g_v^p = 0.0298$ and $g_v^n = -0.5117$ [39]. The axial term $q_A$, is usually smaller by a factor $1/N^2$ and it is 0 for the spin zero nuclei [40] [41] [42]. Thus, the axial term is neglected in our calculation [2]. The weights and abundances of Ge and Si are obtained from the Nuclear Data Center, Japan Atomic Energy Agency to assist the CEvNS cross-section calculation and tabulated into the table below [43].

**Table 1.** *Natural Germanium and natural Silicon with its isotopes and abundance.*

| Ge nuclides | Weight (amu) | Abundance (%) | Si nuclides | Weight (amu) | Abundance (%) |
|---|---|---|---|---|---|
| $^{70}$Ge | 69.92 | 20.57 | $^{28}$Si | 27.98 | 92.223 |
| $^{72}$Ge | 71.92 | 27.45 | $^{29}$Si | 28.98 | 4.685 |
| $^{73}$Ge | 72.92 | 7.75 | $^{30}$Si | 29.97 | 3.092 |
| $^{74}$Ge | 73.92 | 36.50 | | | |
| $^{76}$Ge | 75.92 | 7.73 | | | |

### 3.4 Detector Response Calculation

We defined the detector response reaction rates in terms of incident antineutrino energy as:

$$R(E_v) = N\sigma(E_v)\phi(E_v) \quad (17)$$

where $N$ is the number of nuclides, $\sigma$ is the CEvNS cross-section and $\phi$ is the antineutrino flux. In our study, we also calculate the minimum antineutrino energy required to trigger a 20-eV nuclear recoil for both natural Ge and natural Si, which defined as [35]:

$$E_{min} = \frac{T_R + \sqrt{2MT_R + T_R^2}}{2} \quad (18)$$

### 3.5 Pulse Height Distribution Calculation

Once we obtained the detector response rate from above section, we can use this information to calculate the distribution of the recoil energies deposited in the detector for a given incident antineutrino energy. The calculation of pulse height distribution as a function of recoiled nucleus' energies uses Eq. 19 which provides the probability, $P(E_v \to T_R)$, that given an incident antineutrino with energy, $E_v$, undergoes a CE$\nu$NS interaction, the nucleus will recoil with energy $T_R$. We define the distribution of recoiled energies for a given incident antineutrino energy as below:

$$P(E_v \to T_R) = \frac{d\sigma(E_v, T_R)}{dT_R} / \sigma(E_v) \tag{19}$$

Then, we can define pulse height distribution, $C$ as a function of $T_R$:

$$C(T_R) = \int_{E_v} N\sigma(E_v)\phi(E_v)P(E_v \to T_R)dE_v \tag{20}$$

We assume a 100-kg of natural Ge and natural Si detectors with 20-eV nuclear recoil thresholds are placed 10 m away from the 1 MW TRIGA rector core at Nuclear Science Center, Texas A&M University. Fission parameters and assumptions used to calculate detector response and pulse height distribution are shown in Table 2 [35]:

**Table 2.** *Parameters and assumptions used for detector response and pulse height distribution calculation*.

| **Parameters** | | **Values** |
|---|---|---|
| Fission rate | | 3.1 x $10^{16}$ $s^{-1}$ |
| Antineutrino production rate | | 1.8 x $10^{17}$ $s^{-1}$ |
| Fission Fraction | $^{235}U$ | 0.967 |
| | $^{238}U$ | 0.013 |
| | $^{239}Pu$ | 0.02 |
| | $^{241}Pu$ | 0.001 |
| | $^{238}U$ to $^{239}Pu$ | 0.16 |

## 4. Result and Discussion

A total of 535 fission fragments from $^{235}$U, 575 fission fragments from $^{238}$U, 558 fission fragments from $^{239}$Pu and 595 fission fragments from $^{241}$Pu are selected in our analysis. Many of these fission fragments are common to all four fissionable isotopes. Table 2 shows the number of fission fragments obtained from ENDF/B-VIII.0 and JENDL-2015. A total of 210 isotopes are missing beta intensities in ENDF/B-VIII.0; for these isotopes their continuous beta spectra in Eq. 9 using data from JENDL-2015 will be used, as presented in Appendix section, Table 8.

**Table 3.** *The number of fission fragments data obtained from ENDF/B-VIII.0 library and JEDNL-2015 library.*

| Fissionable Isotopes | ENDF/B-VIII.0 | JENDL-2015 |
|---|---|---|
| $^{235}$U | 393 | 142 |
| $^{238}$U | 385 | 190 |
| $^{239}$Pu | 421 | 137 |
| $^{241}$Pu | 411 | 184 |

In this study, we present four type of corrections to demonstrate the impact of the corrections to the antineutrino spectrum. We will first model the spectrum without considering any isomeric transitions; then we apply the corrections one at a time as shown below:

(a) Including of isomeric transitions using the data variable in ENDF/B-VIII.0;

(b) Improving of isomeric transitions using TAGS data from published literature;

(c) Including of missing isomeric transitions correction with continuous beta spectrum as in Eq. 9 (Gross Theory); and

(d) Correcting for the nucleus' finite size, radiative and weak magnetism

### 4.1 Including Isomeric Transitions

Once a fission product is formed, it may not be formed directly in its ground state. The newly formed nucleus usually stays at a different excited state and de-excites by emitting gamma rays. Thus, it is important to include all transitions ($\beta$ branches) with their beta intensities when computing the antineutrino spectrum. Figure 2 shows the comparison between the antineutrino spectra without considering the transitions (ground state beta intensity = 1) and the spectra including all transitions that are available in the ENDF/B-VIII.0 data library, for $^{235}$U, $^{238}$U, $^{239}$Pu, and $^{241}$Pu.

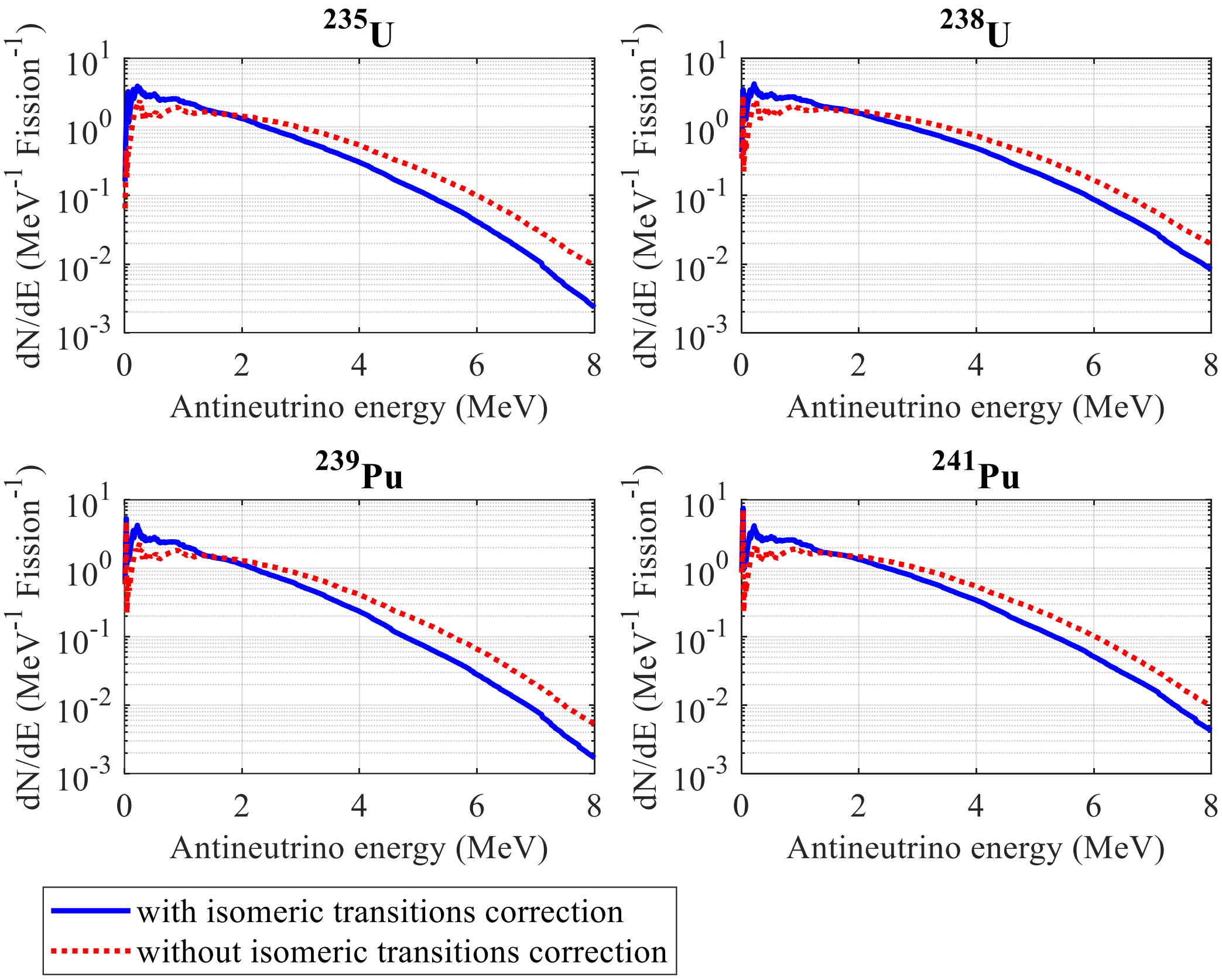

**Fig. 2.** *A comparison between the spectra without isomeric transitions correction and the spectra with all isomeric transitions available in the ENDF/B-VIII.0 data library included, for major fissionable isotopes.*

The inclusion of isomeric states shifts the spectra to lower energies in Fig. 2, as expected, since the excess energy of isomeric states is being rightly discounted. The antineutrino spectra after the isomeric transition corrections have noticeable deviations where the corrected spectra are shifted to lower energies compared to the spectra without the transition corrections in the range 2 to 8 MeV. For $^{235}$U, the discrepancy can be up to a 75.38 % deficit at 8 MeV. This deficit at higher energies in the corrected spectra can also found in $^{238}U$, $^{239}Pu$ and $^{241}Pu$, where the differences are 57.36 %, 67.95 % and 56.23 %, respectively. On the other hand, the spectra without transition corrections tend to underestimate the low energy range (0 - 2 MeV). In the range of 0.5 MeV – 2 MeV, the spectra for all four fissionable isotopes are enhanced 29 % to 37 % on average. In the range of 2 MeV – 8 MeV, the spectra for $^{235}U$, $^{238}U$, $^{239}Pu$, and $^{241}Pu$ shifted to lower energies with an average of 49.35 %, 39.09 %, 47.94 % and 41.07 %, respectively. Thus, inclusion of isomeric transitions shows a significant impact and corrections to the shape of spectra.

### 4.2 Consideration of TAGS Data Set

In our study, we have identified TAGS data sets based on availability found in published literature and included them in our calculations. We included TAGS data sets from Greenwood *et al.* from 1997 to the most recent TAGS data sets by J. Gombass *et al.* (2021) to obtain the antineutrino spectra.  This section is intended to demonstrate the improvement before and after including the TAGS data, once the isomeric states have been included.

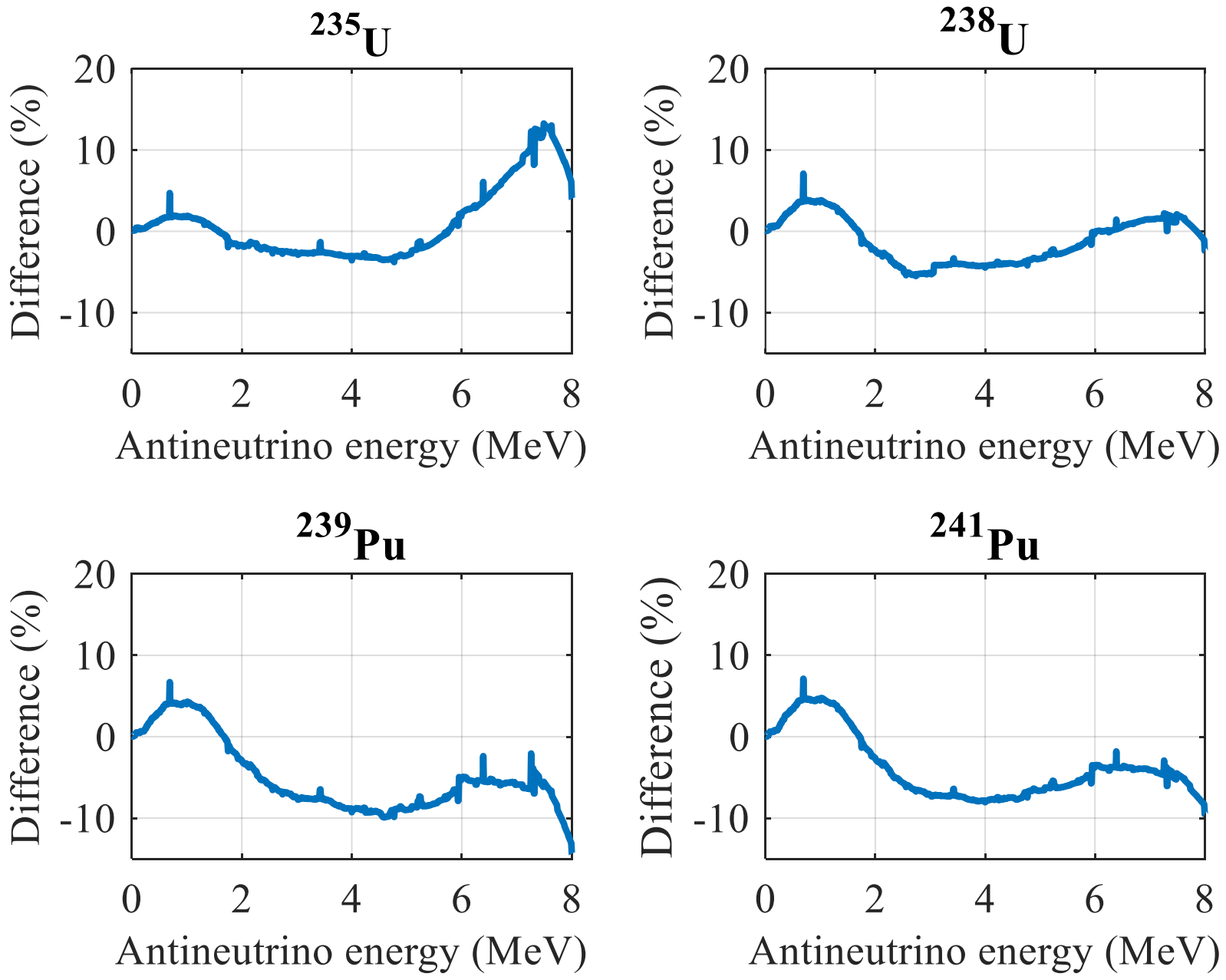

**Fig. 3.** *Percent differences observed in the antineutrino spectra as a function of antineutrino energy, for all four fissionable isotopes, after including the TAGS data set.*

Fig. 3 shows the spectra before and after correcting for the pandemonium effect with the TAGS data sets from the Appendix, Table 7 obtained from the literature review. Overall, there is a noticeable enhancement of the antineutrino spectrum in the energy range of 0 to 2 MeV. For $^{241}$Pu, the deviation is the highest among all four fissionable isotopes, where the discrepancy is 7.13 % in the energy range 0 to 2 MeV whereas, the other spectra increased to 4.71% for $^{235}$U, 7.09 % for $^{238}$U and 6.72 % for $^{239}$Pu. In the range 2 to 6 MeV, the spectra for all four fissionable isotopes decreased by less than 10 %. In the range 6 to 8 MeV, the corrected spectra increased for $^{235}$U and $^{238}$U such that the largest difference observed are 13.25 % and 2.20 %, respectively. The corrected $^{239}$Pu and $^{241}$Pu spectra are decreased in the range of 6 to 8 MeV, where the highest differences are 14.37 % and 9.54 %, respectively. One of the major contributors of this discrepancy is $^{92}$Rb as the difference of their ground state beta intensity between ENDF/B-VIII.0 and TAGS data is 71.57 %. Table 4 shows the different ground-state-Q values quoted in ENDF/B-VIII.0 and TAGS.

**Table 4.** *The difference in beta intensity values between ENDF/B-VIII.0 and TAGS.*

| Nuclide | Q value (MeV) | Ground state beta intensity (%) – ENDF/B-VIII.0 | Ground state beta intensity (%) - TAGS |
|---|---|---|---|
| $^{92}$Rb | 8.095 | $51 \pm 18$ | $87.5 \pm 2.5$ |
| $^{142}$CS | 7.308 | $43 \pm 3$ | $46.4^{+2.7}_{-3.0}$ |
| $^{100}$Nb | 6.396 | $50 \pm 7$ | $46.0^{+8.0}_{-15.0}$ |

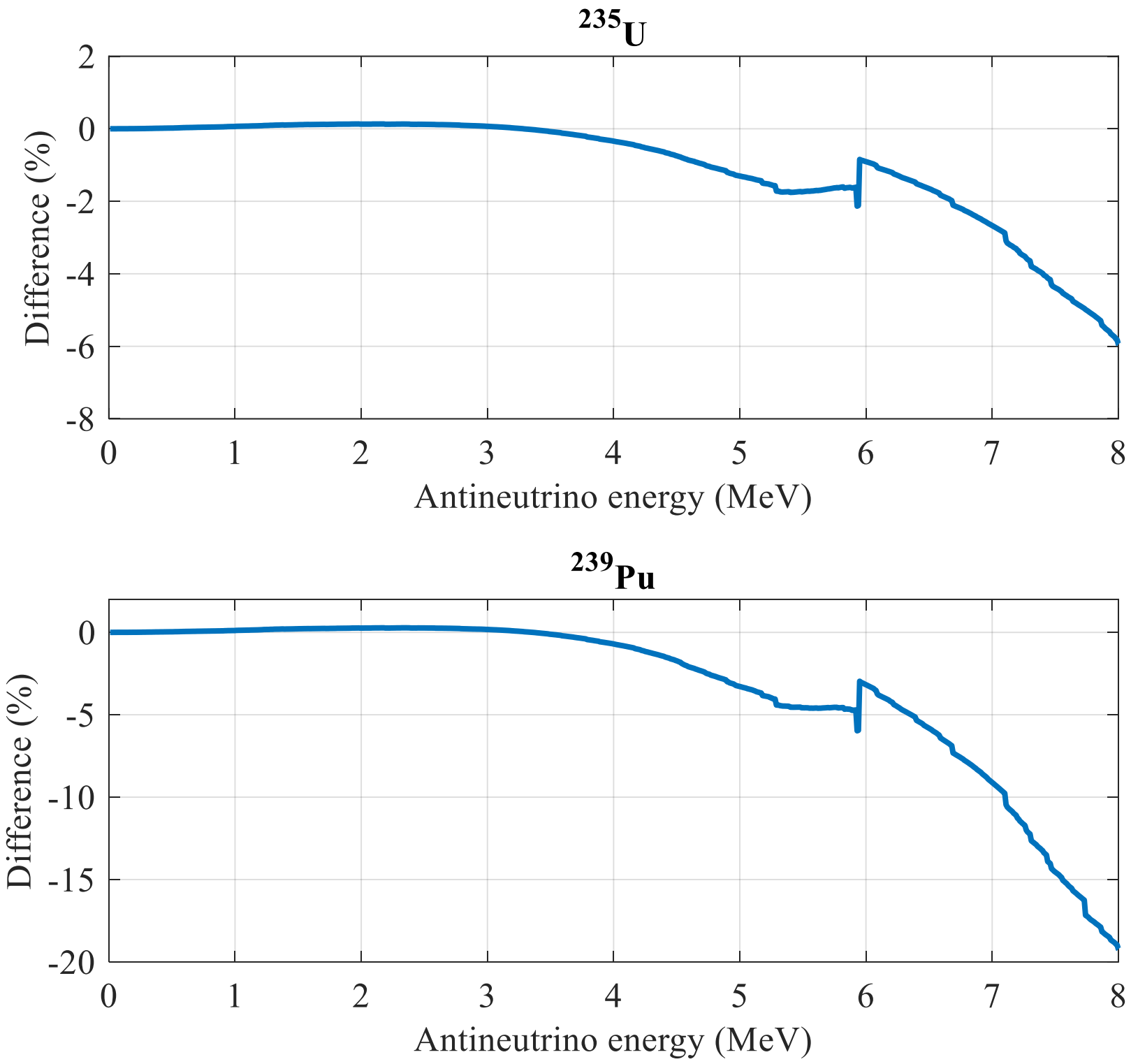

**Fig. 4.** *A comparison between the antineutrino spectra with the most recent TAGS data set by J. Gombass et al. (2021).*

The spectra with the 2021 TAGS data sets have significant impact on energy in the range 6 – 8 MeV. For $^{235}$U and $^{239}$Pu, the average difference with and without the 2021 TAGS data sets in the range 6 – 8 MeV is 3 % and 10 %, respectively. The highest discrepancies can be observed at 8 MeV for both $^{235}$U and $^{239}$Pu, with 5.95 % and 19.22 % deviation, as shown in Fig. 4. The reason for these high discrepancies at high energies is due to the missing isomeric transitions information of $^{103}$Nb and $^{04m}$Nb in ENDF/B-VIII.0 nuclear library.

### 4.3 Missing Isomeric Transitions Data Correction

Some beta intensities and isomeric transitions are not available in ENDF/B-VIII.0, especially for neutron-rich isotopes. Thus, we have used the predicted beta continuous spectrum in Eq. 9 from JENDL-2015 and integrated them into our study. Figure 5 below shows the difference of the spectra using data from JENDL-2015 and the spectra without using the continuous spectrum (assuming the beta intensity to the ground state is 100 %).

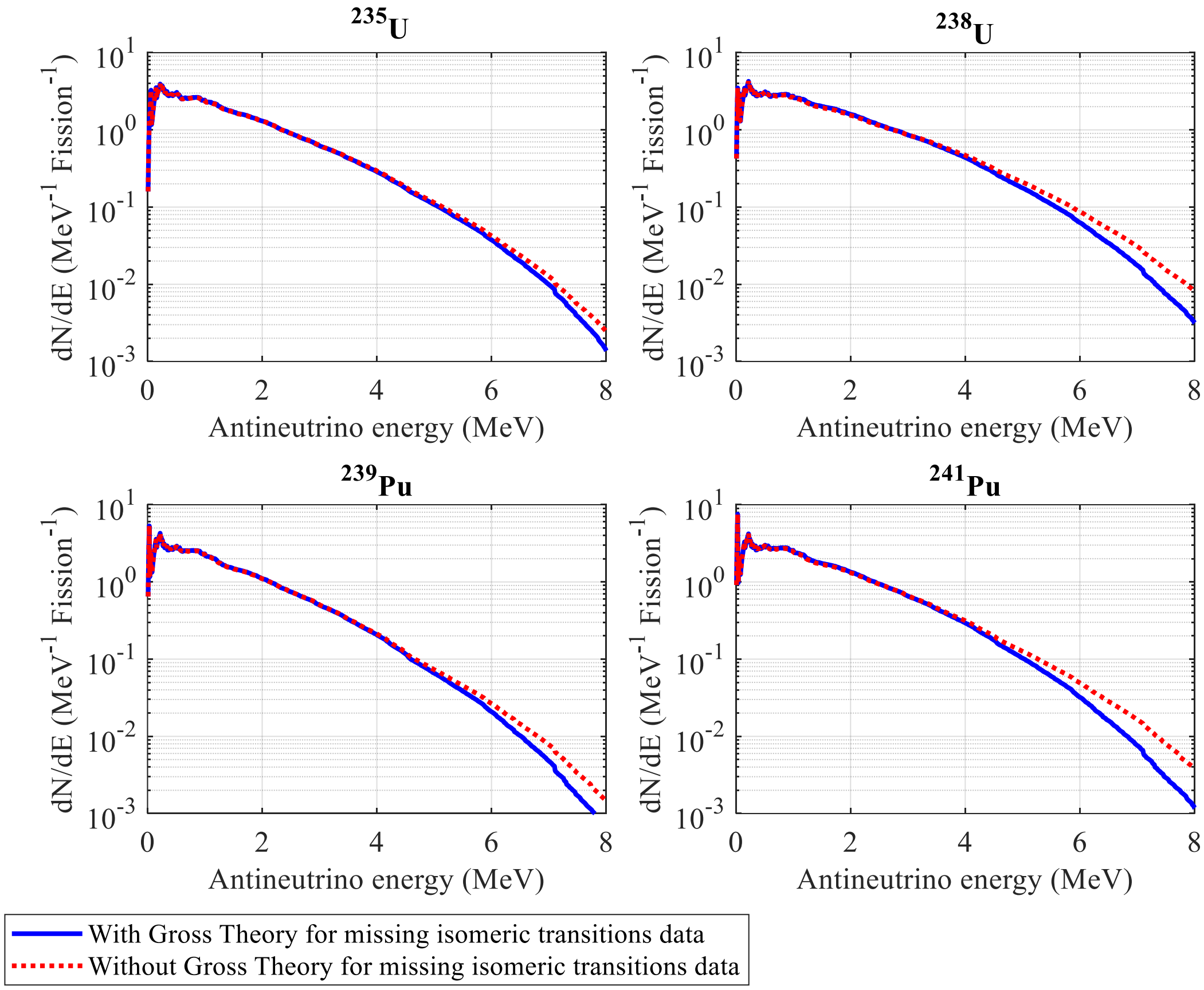

**Fig. 5.** *A comparison of the antineutrino energy spectra with the neutron-rich fission products correction, and the spectra without the correction. The corrected antineutrino spectra reduce the counts at higher energies.*

For $^{235}$U, the spectrum without neutron-rich fission products (FPs) correction is reduced by an average of 4.43 % in the range of 3 MeV to 6 MeV. This discrepancy gets larger from 6 MeV to 8 MeV, where the spectrum without the correction is reduced by 11.56 % to 43.56 % after considering the continuous spectrum from JENDL-2015. The $^{238}$U spectrum without the correction is also reduced by an average of 11.69 % from 3 MeV to 6 MeV. The discrepancy is even higher where the spectrum before the correction is reduced by 27.78 % to 61.74 % within 6 – 8 MeV. The uncorrected $^{239}$Pu and $^{241}$Pu spectra are decreased by 7.57 % and 13.6 % respectively, after the missing isomeric transitions data correction for neutron rich isotopes in the range of 3 MeV to 6 MeV. However, between 6 and 8 MeV, the discrepancy of the corrected and uncorrected $^{241}$Pu is the highest among all four fissionable isotopes, where the uncorrected spectrum is reduced by 34.76 % to 69.46 %.

### 4.4 Finite Size, Radiative and Weak Magnetism Corrections

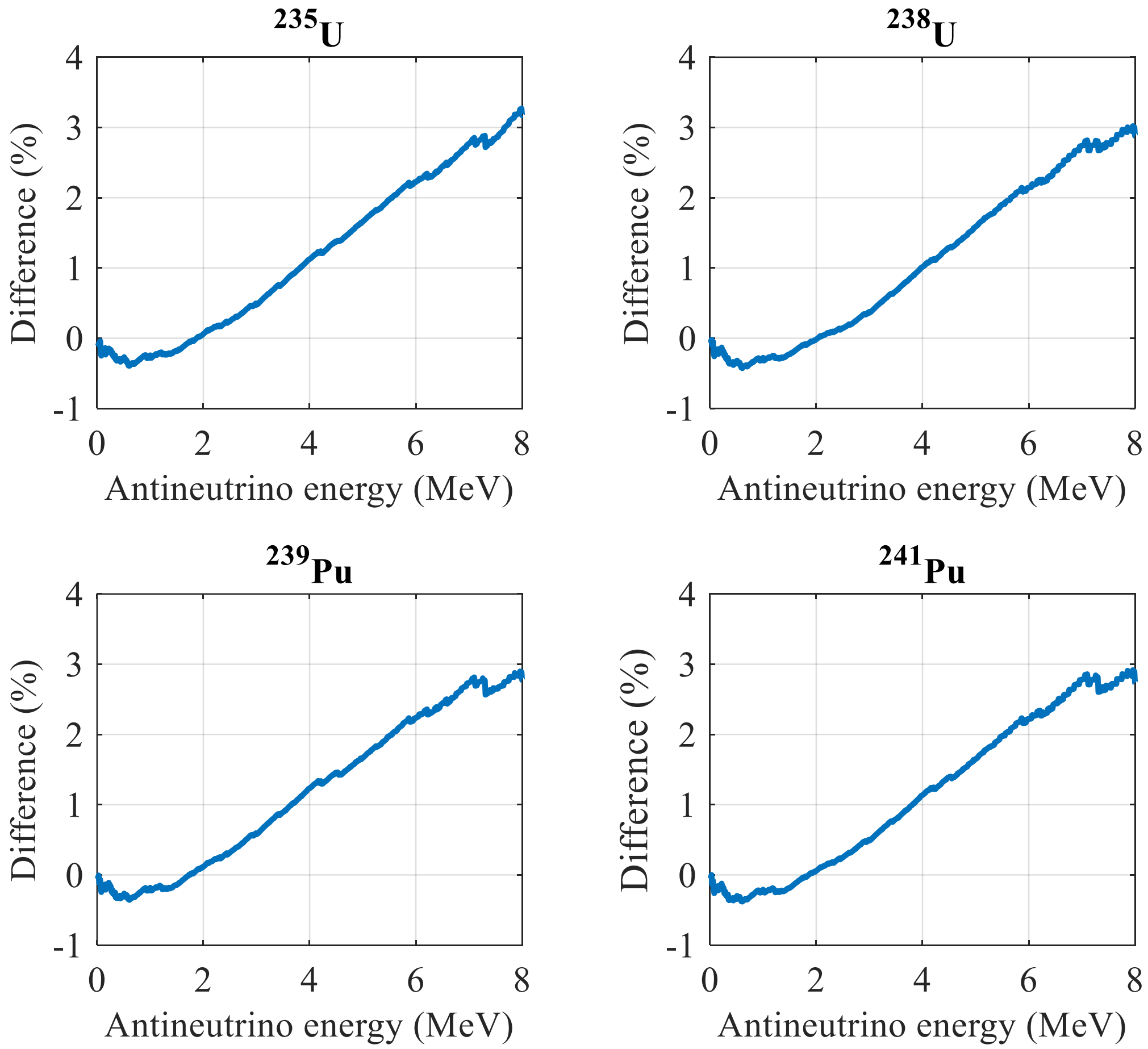

**Fig. 6.** *Percent difference of the corrected antineutrino spectra as a function of the antineutrino energy. Corrections were made for finite size, weak magnetism and radiative corrections. The corrected spectrum sees an increase in counts towards higher energies.*

Next, we applied finite size, radiative and weak magnetism corrections on the result from section 3.1 given by Eq. 5 - 8. The differences of the spectrum with and without finite size, weak magnetism and radiative are shown in Fig 6. In the energy range of 0 to 2 MeV, the differences between corrected and uncorrected spectrum for all four fissionable isotopes are no more than 0.43 % smaller. From 2 MeV, the difference between corrected and uncorrected spectrum for all four fissionable isotopes start to increase gradually to less than 3.27 % at 8 MeV. In Fig. 6, we find that the finite size, weak magnetism and radiative corrections have greater impact on the spectrum in the high energy range. The finite size correction is the main contributor where 2.04 % to 2.20 % average enhancement can be observed from all four fissionable isotopes from 2 – 8 MeV. In the energy range 0 – 2 MeV, only 0.22 % and 0.31 % average deficit are observed with finite size correction. For weak magnetism correction, 0.61 % to 0.71 % average deficit deviations are found after the correction for all four fissionable isotopes at 2 – 8 MeV. On the other hand, 0.06 % to 0.08 % enhancement are observed in the energy range 0 to 2 MeV. Lastly, radiative correction contributes about 0.20 % to 0.24 % increment from 2 – 8 MeV for $^{235}$U, $^{238}$U, $^{239}$Pu and $^{241}$Pu. However, from 0 to 2 MeV, there are 0.02 % to 0.03 % average deficit with radiative correction.

**4.5 Comparison Corrected Spectrum with Previous Works**

To establish the validity of our method, we compared our modeled spectra with previous works from Hayes *et al.* (summation method) and Huber *et al.* (conversion method) in this section. It is worth mentioning, that Ishimoto *et al.* (2002) developed an estimation method without including all transitions from the fission fragments [44]. By introducing a correction factor, $F_a(E_v) = C(1 - \frac{E_v}{Q})^n$ into Eq. 3 which alters the spectrum such that it matches with the others experimental spectra. The notation $C$ is the normalization factor and $n$ is an adjustable parameter. Our previous preliminary study was performed using this method to model the antineutrino spectra, and the results can be found from the reference [4]. We also present our result using Ishimoto's method here for comparison purposes.

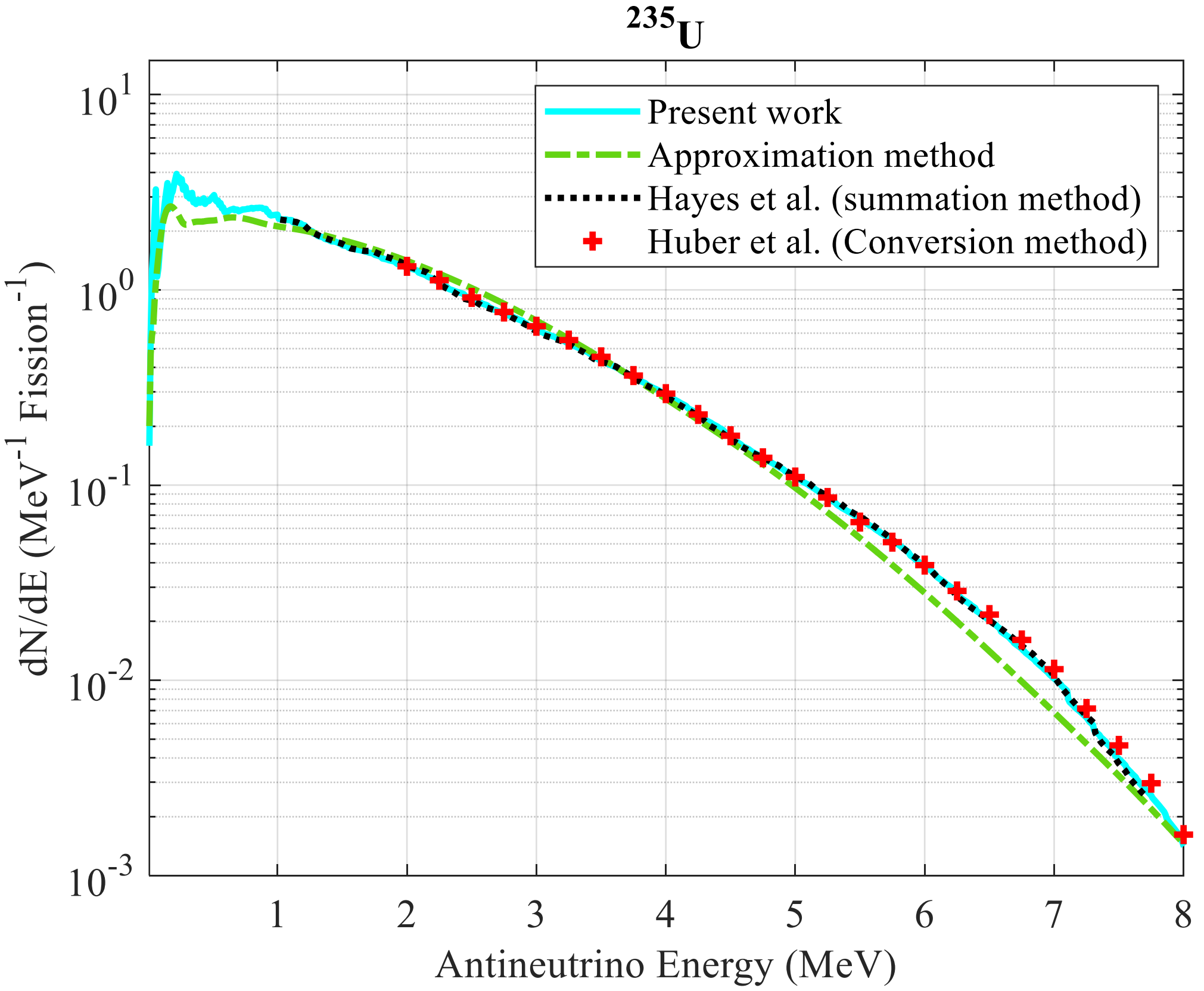

**Fig. 7.** *The $^{235}$U spectra from this study (summation and approximation method), Hayes et al. (summation method) and Huber et al. (conversion method).*

Fig. 7 shows the comparison of calculated $^{235}$U antineutrino spectrum with Huber *et al.* (2011) [11] and Hayes *et al.* (2016) spectrum, which used the conversion and summation method, respectively [11] [8]. Our calculated $^{235}$U antineutrino spectrum agrees with Huber *et al.* spectrum with an average 2.07 % discrepancy in the range of 2 – 6 MeV. The discrepancy is higher between 6 and 8 MeV, where 3 to 18 % deviation is observed. Our calculated spectrum has lower discrepancies when compared to spectra by Hayes *et al.*, where an average of 1.28 % is calculated in the energy range 1 – 7.5 MeV. The $^{235}U$ spectrum using the Ishimoto's approximation method underestimates the spectrum at the energy range lower than 1.5 MeV with an average of 16.78 % difference compared to the present work. The approximated $^{235}U$ spectrum has better agreement with our $^{235}U$ spectrum using summation method in the range of 4 and 5 MeV with an average of 7 % difference. The deviation of the approximation methods becomes greater in the range 5 - 8 MeV where 11 % to 33 % is observed.

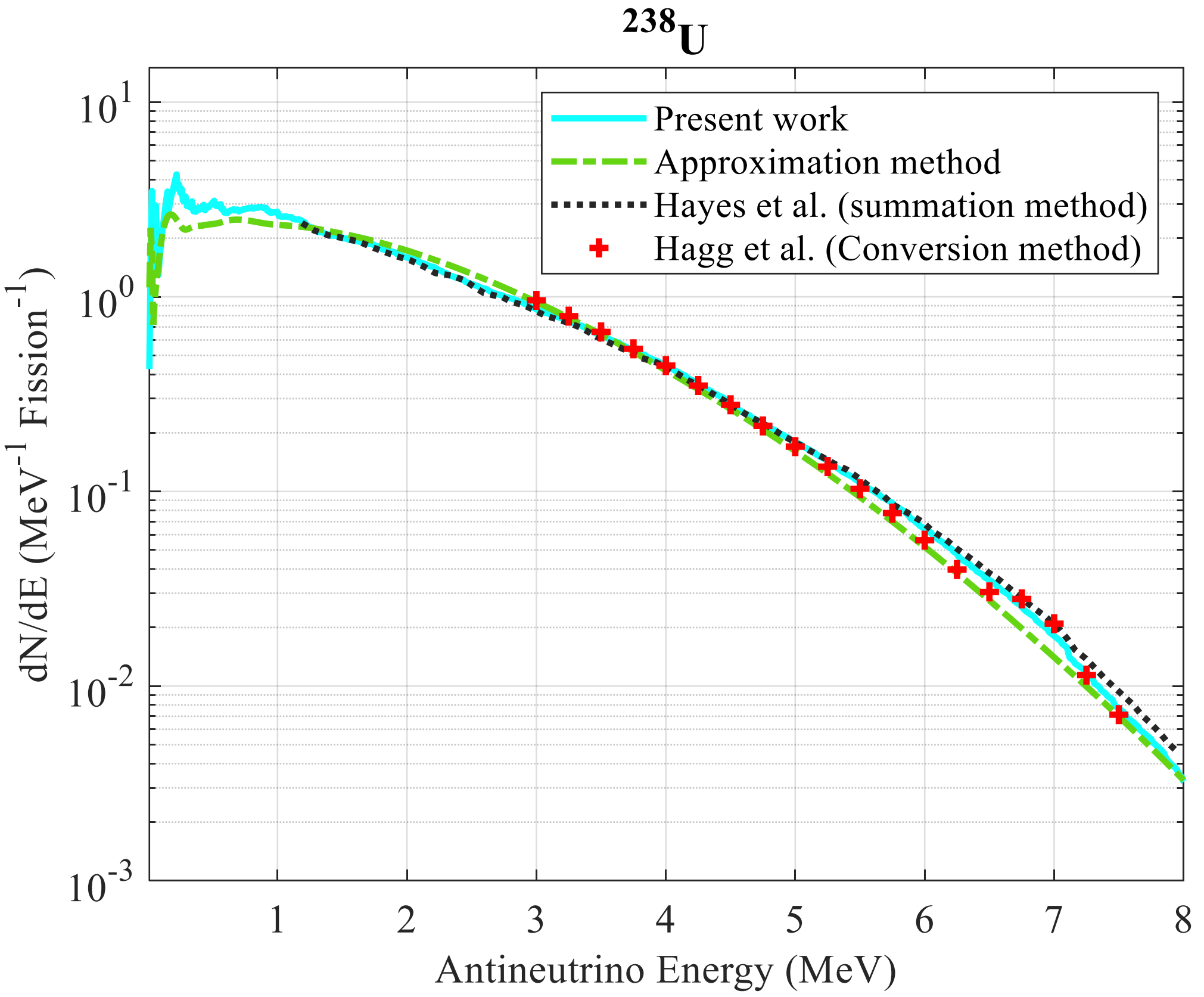

**Fig. 8.** *The $^{238}U$ spectra from this study (summation and approximation method), Hayes et al. (summation method) and Haag et al. (conversion method).*

Overall, the corrected $^{238}U$ spectrum has lower discrepancies when compared to Hayes *et al.* (2016) conversion result than Haag *et al.* (2013) spectrum [12]. From 1 to 6 MeV, the discrepancy with Hayes *et al.* spectrum is no more than 3.6 %. However, from 6 MeV to 8 MeV, the deviation gradually increases from 6 % to 30 %. On the other hand, our calculated results have larger deviations from Haag *et al.* (2013) results from 3 to 7 MeV where there is an average of 7.51 % difference. Beyond 7 MeV, our results have better agreement with Hagg *et al.* results with no more than 7 % difference. Overall, the $^{238}U$ spectrum using Ishimoto's method underestimates the spectrum by an average of 16.34 % difference at low energies (< 1.5 MeV). The difference is greater from 5 MeV to 8 MeV where 10 % to 24 % deviation is observed.

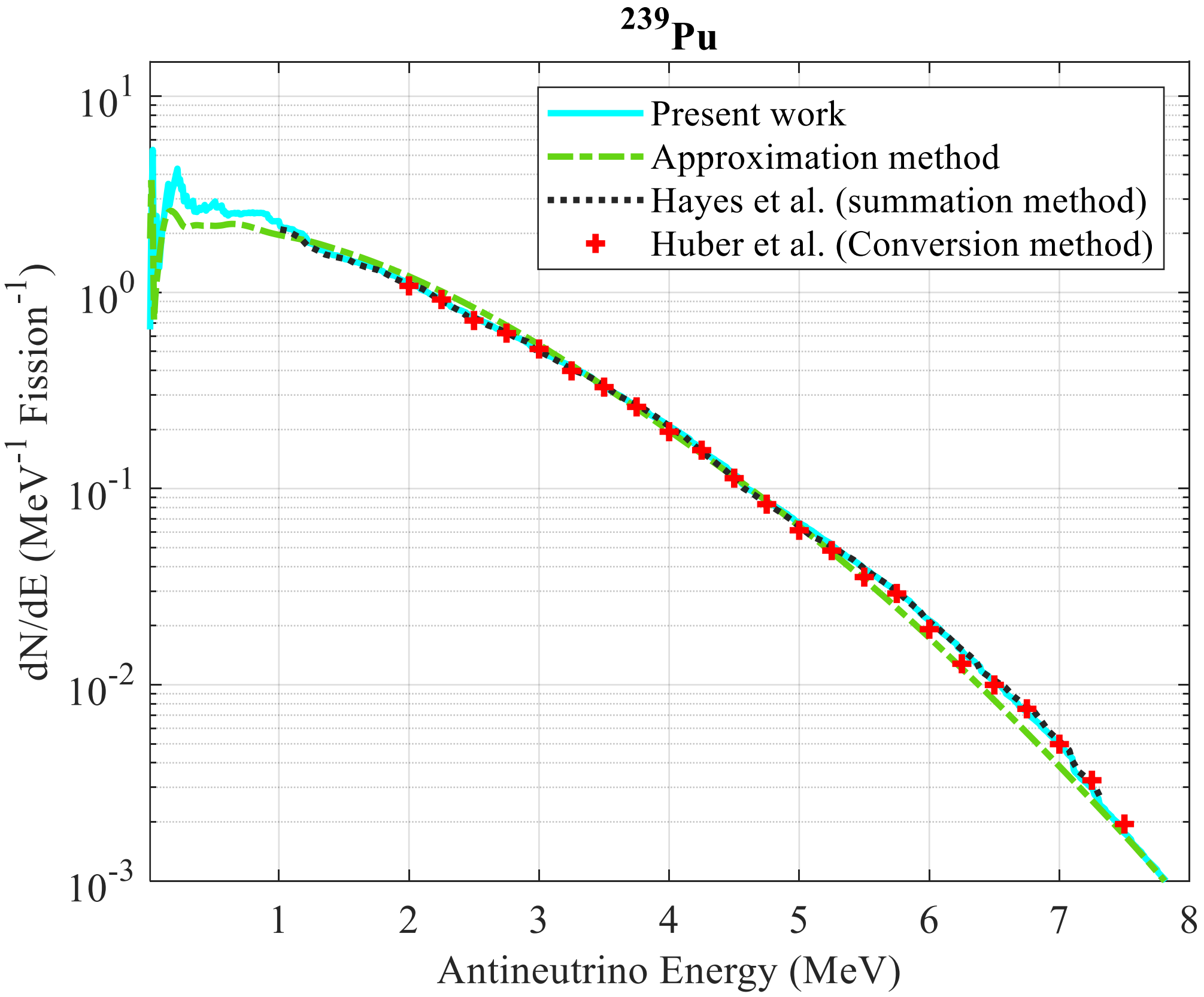

**Fig. 9.** *The $^{239}Pu$ spectra from this study (summation and approximation method), Hayes et al. (summation method) and Huber et al. (conversion method).*

From 2 – 8 MeV, the spectrum by Huber *et al.* spectrum has an average deviation of 5.96 % with our calculated $^{239}Pu$ results, where the highest deviation is at 7.75 MeV with 25.81 % difference. Hayes *et al.* spectrum has an average difference of 2.17 % from our results in the range 1 MeV – 7.25 MeV. The decrepanies between our results and Hayes *et al.* are no more than 3.4 % from 1.25 MeV to 6.5 MeV. For $^{239}Pu$ spectrum using the Ishimoto's approximation method, the spectrum is underestimated with an average of 21.22 % compared to our spectrum using summation method. In the energy range 4 to 8 MeV, the approximated spectrum is underestimated with the highest deviation of 22 % difference at 7 MeV.

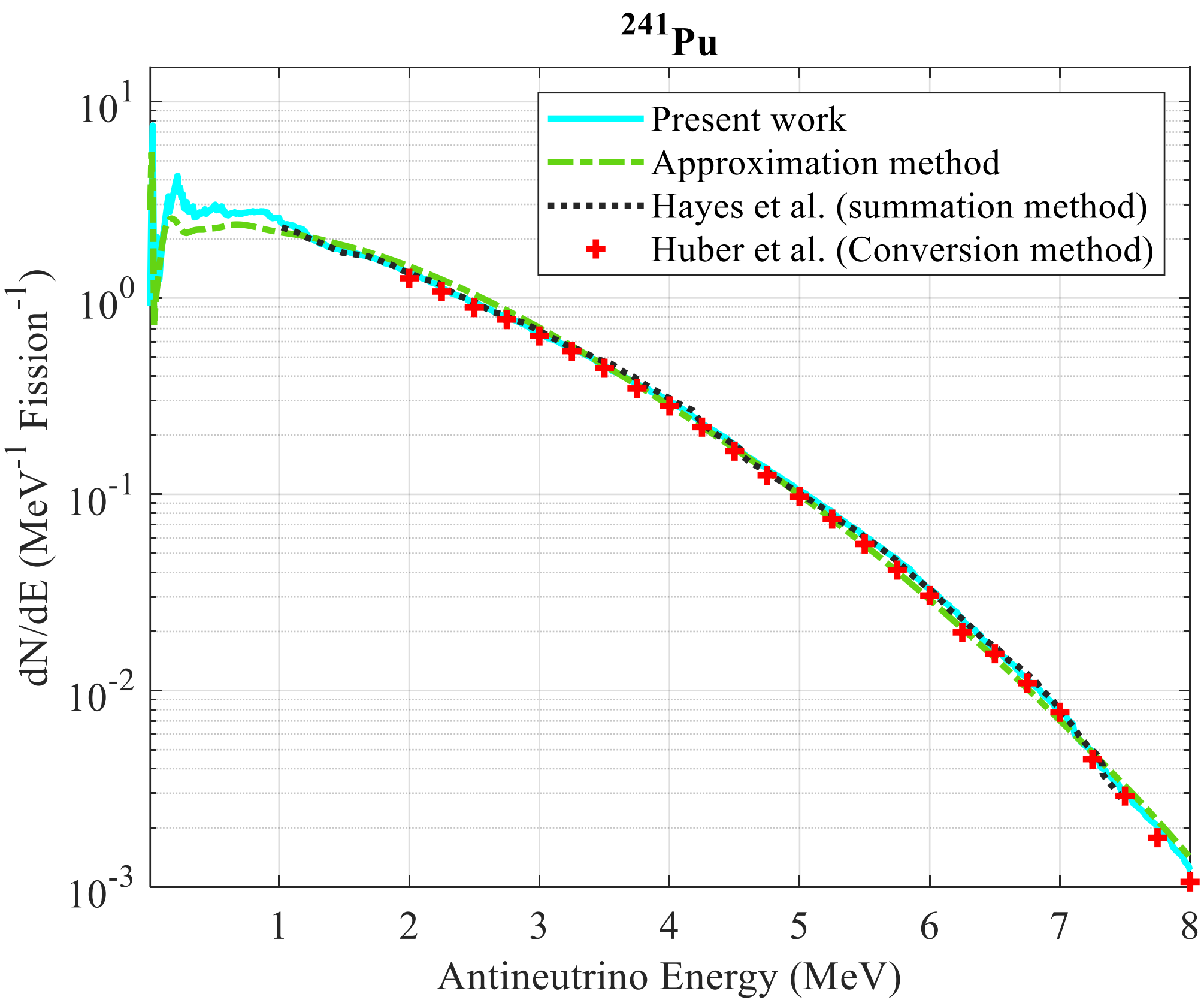

**Fig. 10.** *The $^{241}Pu$ spectra from this study (summation and approximation method), Hayes et al. (summation method) and Huber et al. (conversion method).*

Our calculated $^{241}$Pu spectrum has larger discrepancies with Huber *et al.* result than with Hayes *et al.* spectrum. Overall, the calculated spectrum has an average discrepancy of 6.61 % with Huber *et al.* spectrum from 2 – 8 MeV, where the highest difference is observed at 6.25 MeV with a 14.88 % deviation. For Hayes *et al*., our results agreed with no more than 3.5 % difference, except at 1 MeV with 6.59 % and 7.5 MeV with 14.24 %. Overall, the average discrepancy between our corrected spectrum and Hayes *et al.* spectrum is about 2.01 %. The result of the spectrum using the Ishimoto approximation method is underestimated at low energy range less than 1 MeV with an average of 20.84 % deviation.

### 4.6 Detector Response and Pulse Shape Distribution

In this section, we would like to demonstrate the impact of various corrections to the detector response. We also present pulse shape distribution with our corrected spectra using the results from the detector response.

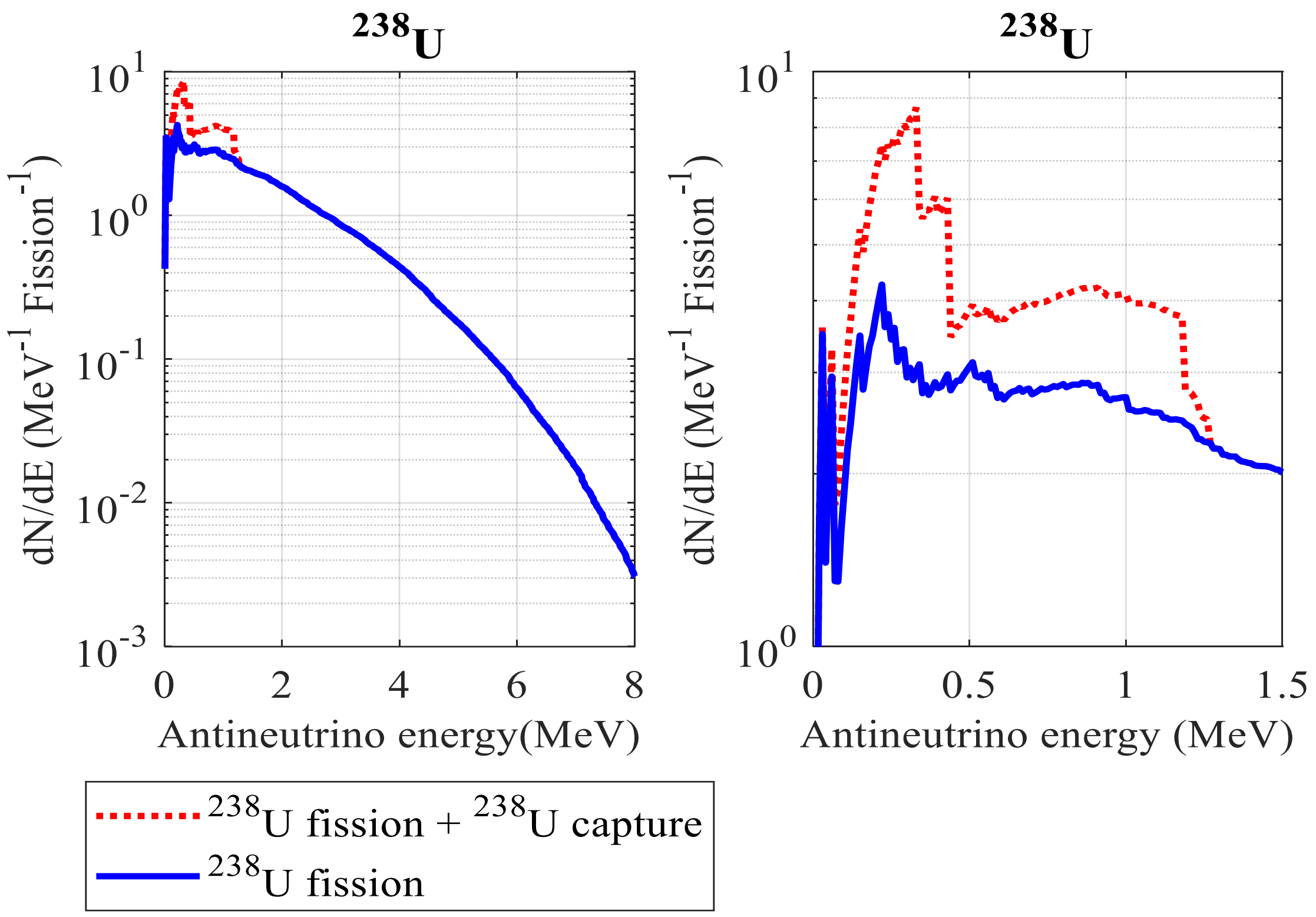

**Fig. 11.** *The antineutrino spectrum from $^{238}U$ fission event and the $^{238}U$ capture event. The breeding information due to capture lies below the 1.3 MeV.*

The contribution $^{238}$U capture event to antineutrino production should be taken into consideration when calculating the detector response with CE$\nu$NS because it has the potential to measure these energies. In 1-MW TRIGA reactor, the capture event of $^{238}$U contributes about 0.16 of fission fraction. In a typical power reactor, the fraction of $^{238}$U capture events is even higher with 0.6 [35]. Figure 11 shows the antineutrino spectrum of $^{238}$U with and without neutron capture events. A $^{238}$U neutron capture event will produce two antineutrinos from two beta decay of $^{239}$U with an average energy of 0.54 MeV, each. Table 5 shows the average number of antineutrino emitted and its average energies from our corrected spectra.

**Table 5.** *The average number of antineutrino emission $N_\nu$ and its average energies ($E_\nu$) for $^{235}U$, $^{238}U$, $^{239}Pu$ and $^{241}Pu$.*

| Fissionable Isotopes | $E_\nu$ (MeV) | $N_\nu$ (per fission) |
|---|---|---|
| $^{235}$U | 1.49 | 6.08 |
| $^{238}$U | 1.65 | 7.17 |
| $^{239}$Pu | 1.36 | 5.51 |
| $^{241}$Pu | 1.48 | 6.26 |
| $^{238}$U to $^{239}$Pu | 0.54 | 2.00 |

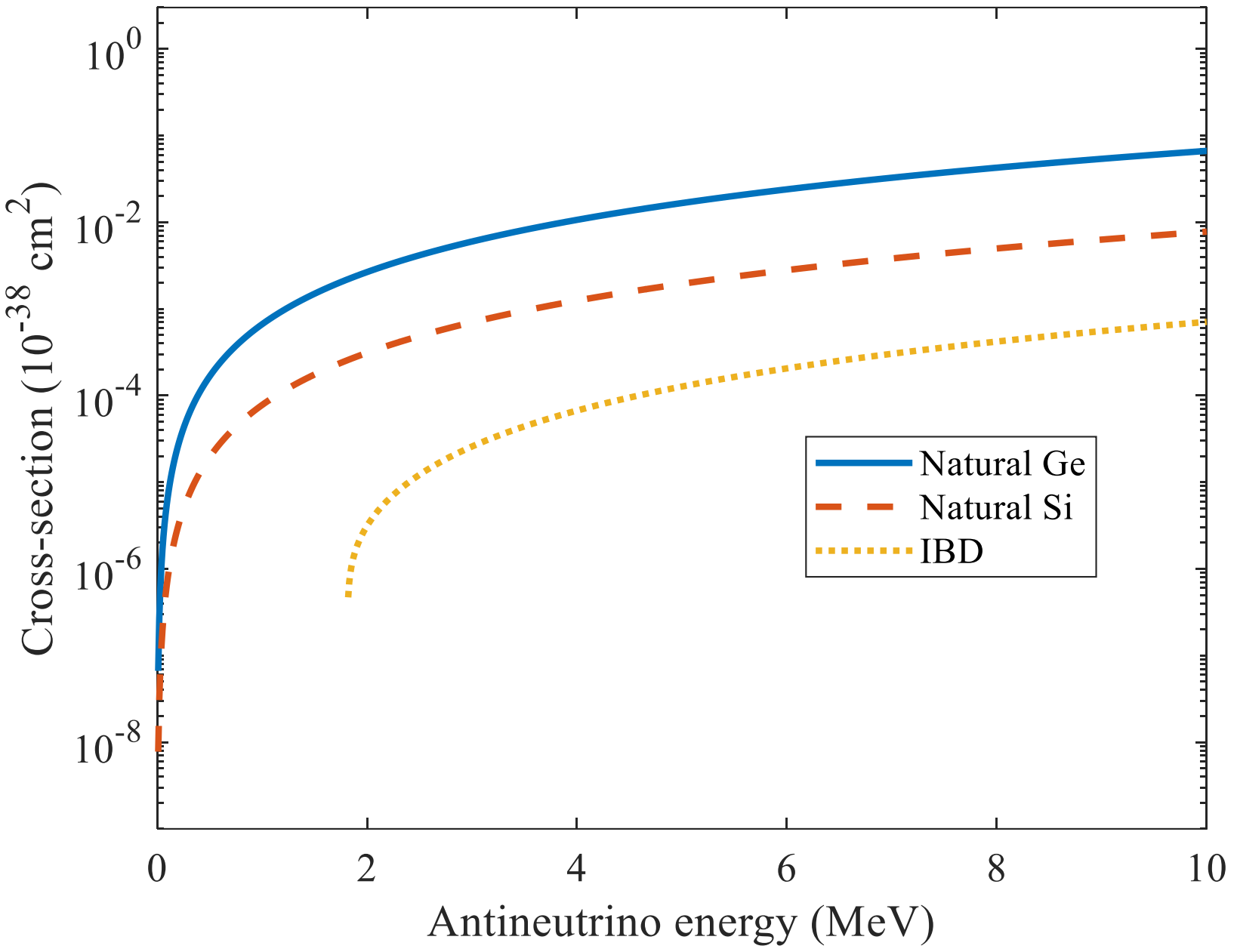

**Fig. 12.** *CEvNS cross-section as a function energy for natural Ge, natural Si, and IBD.*

Fig. 12 shows the CEvNS cross-section of natural Ge and natural Si as a function of incident antineutrino energy, calculated using Eq. 11. For comparison purposes, the IBD cross-section is also included in Fig. 10. There is no energy intrinsic threshold of CEvNS cross-section for natural Ge and natural Si, its threshold is determined by the sensor. In contrast, the IBD cross-section has an intrinsic threshold of 1.806 MeV to initiate the reaction. Overall, the flux weighted average CEνNS cross-sections for natural Ge and natural Si are $4.35 \times 10^{-40} cm^2$ and $5.09 \times 10^{-41} cm^2$, respectively. The IBD flux weighted average cross-section is $4.56 \times 10^{-42} cm^2$. In our study, we assume that the detectors are placed 10 m away from the outer core of the 1-MW TRIGA reactor and the detectors weight 100 kg. Additionally, the CEνNS detectors have a detection threshold of 20-eV nuclear recoil. This corresponds to 0.82 MeV incident antineutrino energy in natural Ge and 0.51 MeV incident antineutrino energy in natural Si. The table below shows comparison between the difference before and after the various corrections:

**Table 6.** *The detector response of natural Ge and natural Si detectors in events/day with various type of corrections implemented sequentially, one after another.*

| **Types of correction** | **Detector response (Events/day)** | |
|---|---|---|
| | **Natural Ge** | **Natural Si** |
| Case 1<br>a) ENDF/VIII.0 data<br>- No transition is considered | 73.49 | 13.13 |
| Case 2<br>a) Isomeric transitions<br>- ENDF/VIII.0 isomeric transitions | 46.07 | 8.31 |
| Case 3<br>a) Isomeric transitions<br>- ENDF/VIII.0 isomeric transitions<br>- TAGS corrected isomeric transitions | 45.33 | 8.18 |
| Case 4<br>a) Isomeric transitions<br>- ENDF/VIII.0 isomeric transitions<br>- TAGS corrected isomeric transitions<br>- Gross Theory for missing isomeric transitions data | 43.92 | 7.93 |
| Case 5<br>a) Isomeric transitions<br>- ENDF/VIII.0 isomeric transitions<br>- TAGS corrected isomeric transitions<br>- Gross Theory for missing isomeric transitions data | 44.25 | 7.99 |
| b) Finite size, radiative and weak magnetism correction | | |
| <u>Case 6</u><br>a) ENDF/VIII.0 data<br>- No transition is considered<br>- Ishimoto's approximation method | 44.14 | 7.96 |

Natural Ge and natural Si detectors response rate after all corrections (including $^{238}$U neutron captures) are 44.25 events/day and 7.99 events/day. The biggest impact on the detector responses is due to the inclusion of the isomeric transitions from ENDF/B-VIII.0. The difference in the detector response with and without isomeric transitions consideration in both natural Ge and natural Si is about 37 %. The finite size, radiative, weak magnetism corrections are the least contributors, where only a 0.8 % difference is observed in the detector response when accounting for these factors. Next, our results show that taking TAGS data set into consideration contribute a 1.6 % deviation in the detector response. Additionally, the detector response for natural Ge and natural Si without J. Gombass *et al.* TAGS data sets are 45.52 events/day and 8.21 events/day, respectively. The difference in the detector response for natural Ge and natural Si with and without J. Gombass *et al.* TAGS data sets are 0.42 % and 0.37 % deviation. The spectra with isomeric transitions correction using the Gross Theory contributes to a 3.1% increment in the detector response.

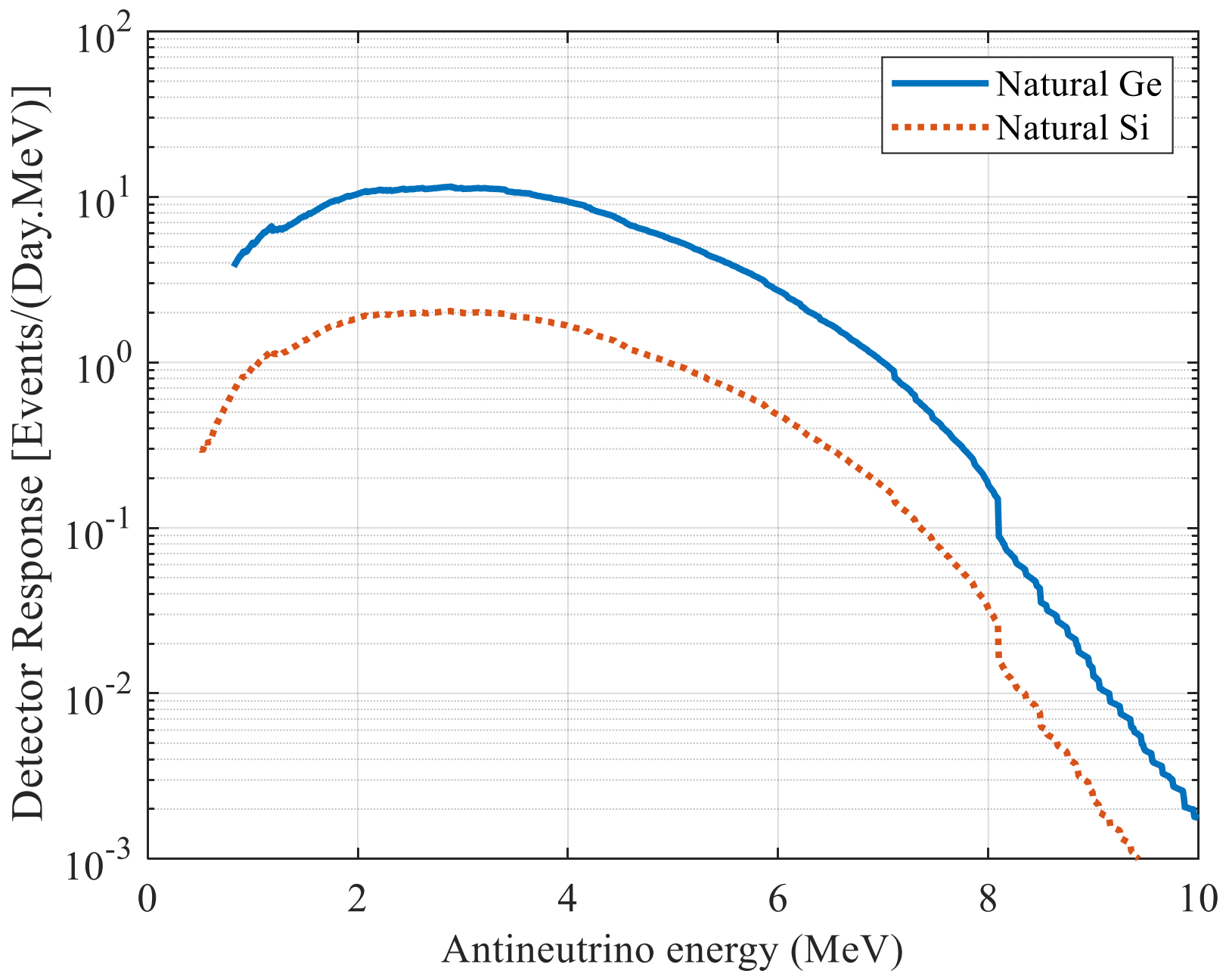

**Fig. 13.** *Detector response as a function antineutrino energy for natural Ge and natural Si with a threshold (20-eV nuclear recoil), for a 100-kg detector, 10 m from the core.*

Figure 13 shows the detector response as a function antineutrino energy for natural Ge and natural Si with an energy threshold of 20 eV nuclear recoil energy. The detector response was calculated using corrected spectra and it yields 44.25 events/day for natural Ge and 7.99 events/day for natural Si. Overall, the detector response of natural Ge detector is higher than natural Si detector by a factor 5.5. However, the natural Si detector is more sensitive at lower energies where it can detect anything above 0.51 MeV, corresponding to a 20-eV recoil energy threshold compared to 0.82 MeV of natural Ge detector.

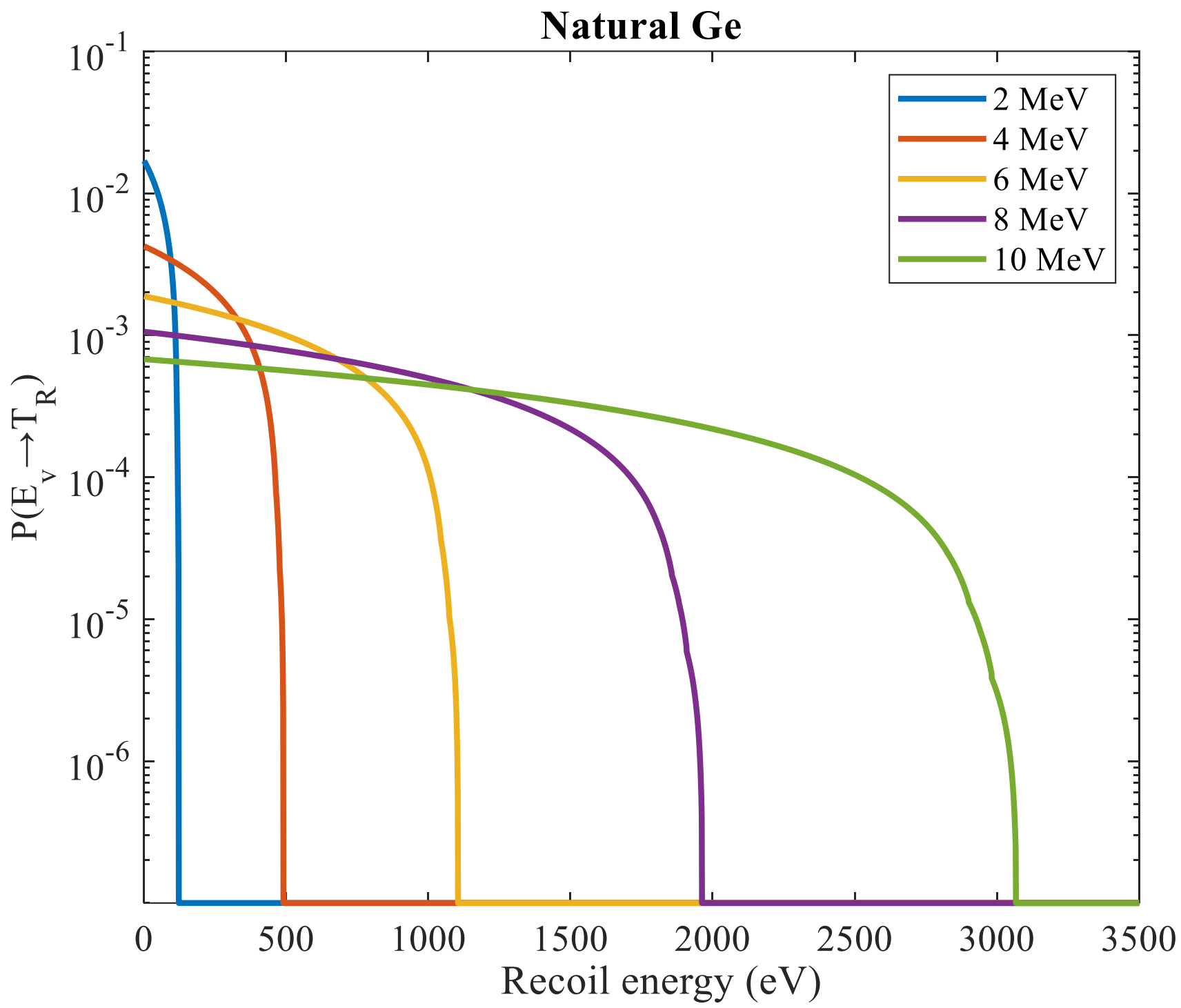

**Fig. 14.** *The probability distribution of nuclear recoil energies for a given incident antineutrino energy in a natural Ge detector*.

Figure 14 shows the probability distribution of nuclear recoil energies for a given incident antineutrino energy with a natural Ge detector. Figure 14 also shows the maximum recoil energies that can be limited by the incident antineutrino energy and the type of materials. The maximum recoil energies that can be produced in a natural Ge detector for 2 MeV, 4 MeV, 6 MeV, 8 MeV and 10 MeV are 123 eV, 491 eV, 1104 eV, 1963 eV and 3066 eV, respectively. These values will change with different incident antineutrino energies and different types of semiconductor materials.

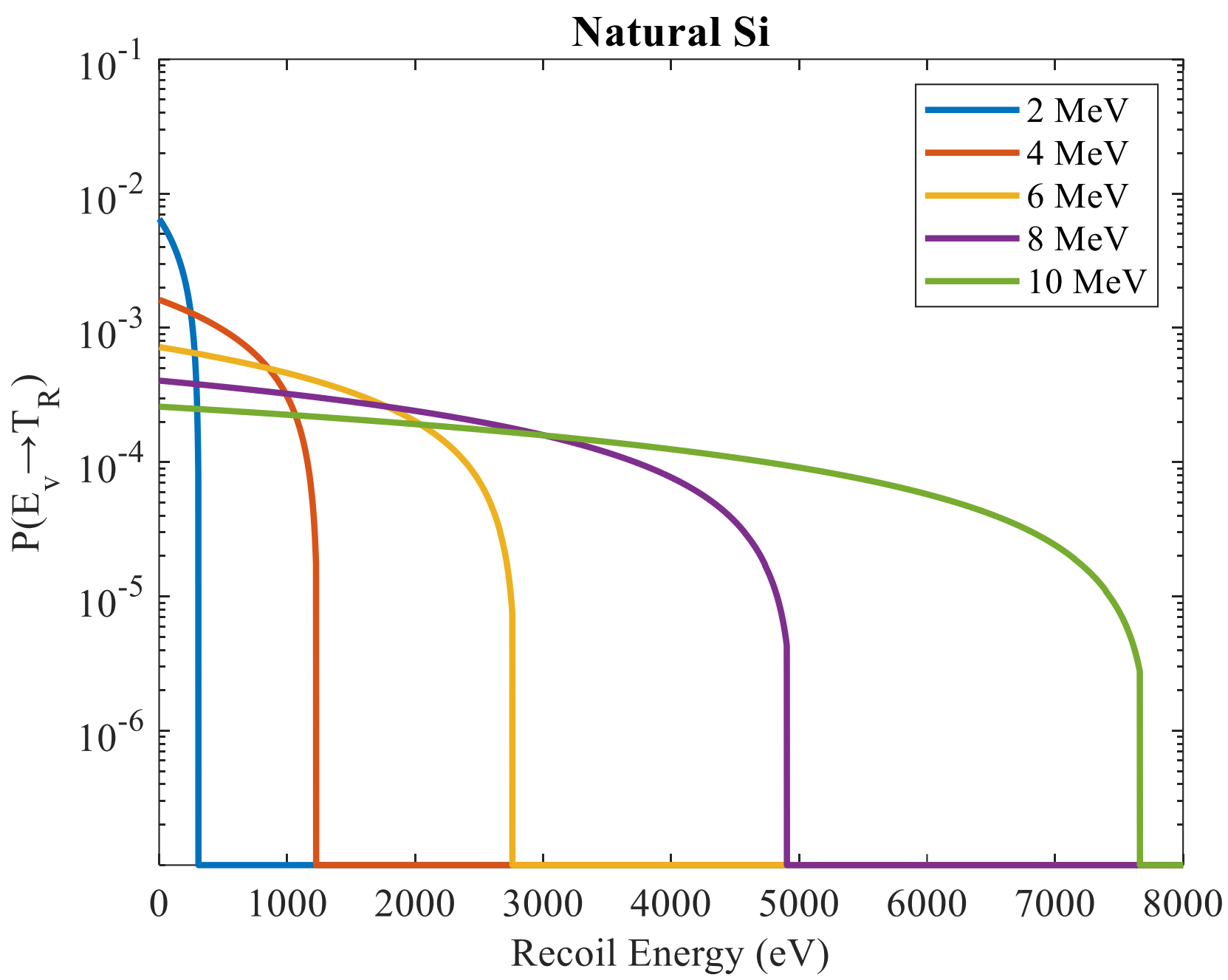

**Fig. 15.** *The probability distribution of nuclear recoil energies for a given incident antineutrino energy in a natural Si detector*.

For natural Si, Fig. 15 shows the probability of nuclear recoil energies for a given incident antineutrino energy has the same trend as natural Ge, where the probability of recoil energy decreases at higher recoil energy. However, the maximum recoil energy of natural Si is higher because the mass of Si is lighter than natural Ge, as in Eq. 12. The maximum recoil energies that can be produced in a natural Si detector for 2 MeV, 4 MeV, 6 MeV, 8 MeV and 10 MeV are 307 eV, 1227 eV, 2759 eV, 4905 eV and 7662 eV, respectively. These values will change with different incident antineutrino energies and different types of semiconductor materials.

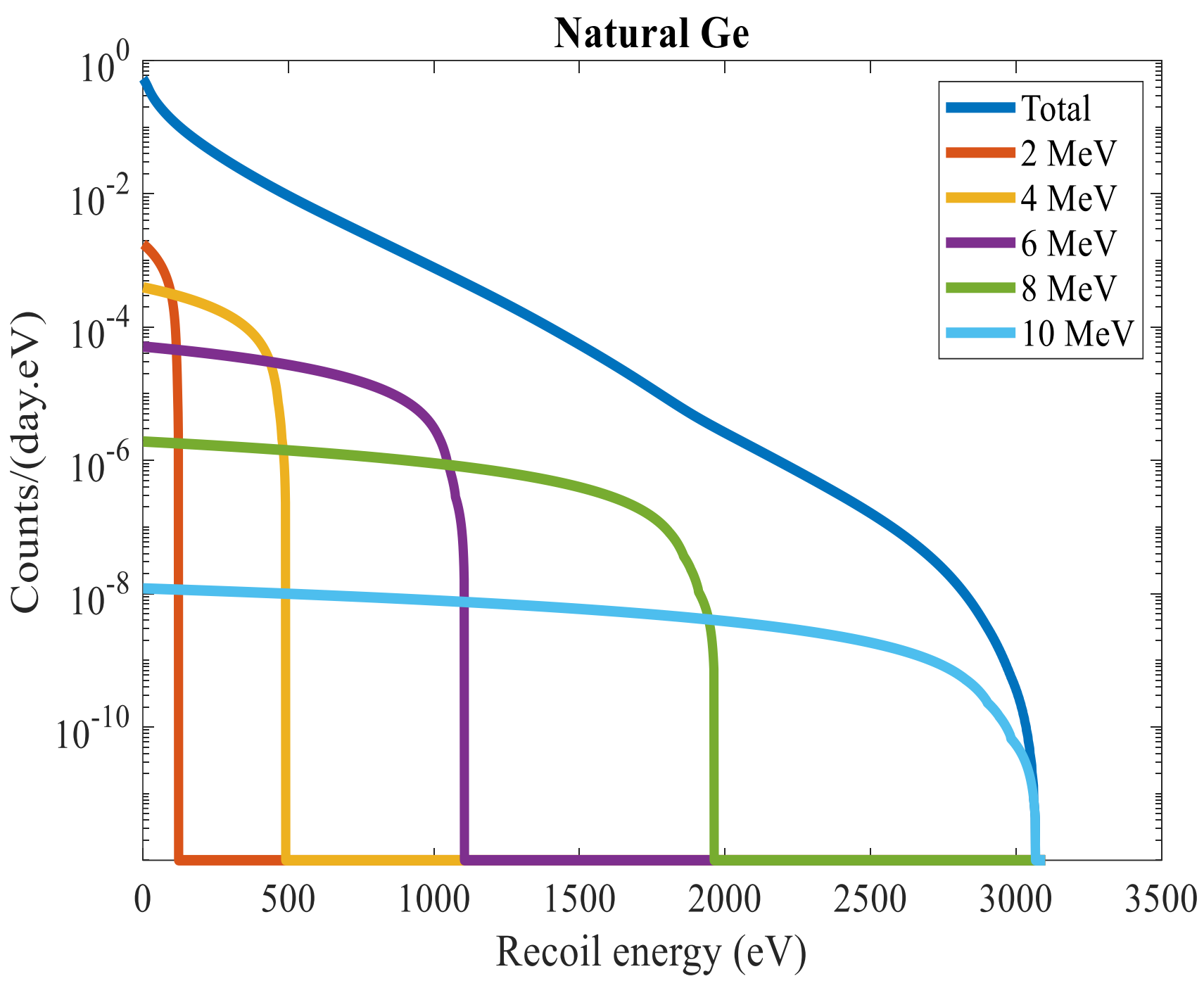

**Fig. 16.** *The calculated pulse height distribution from a natural Ge detector*.

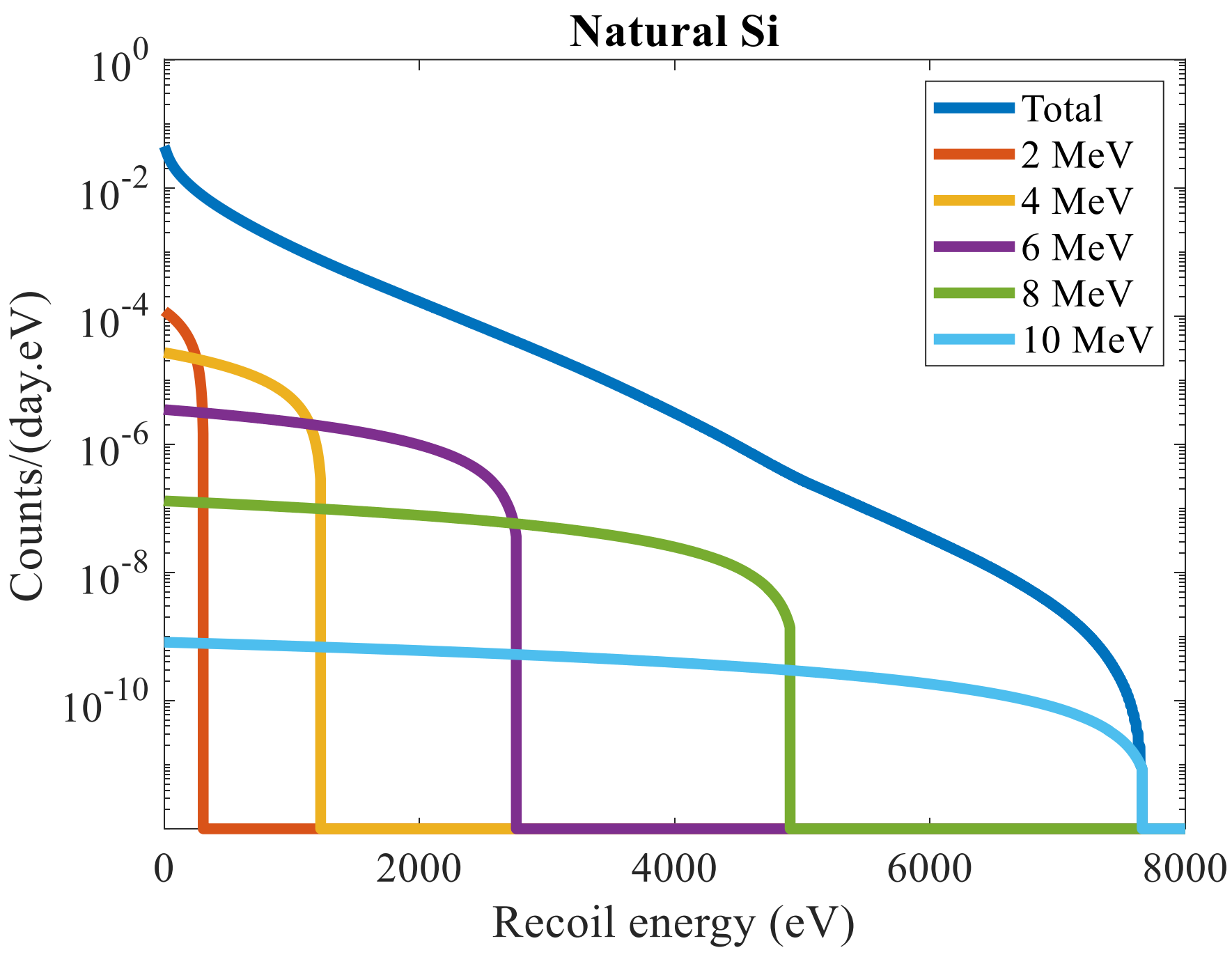

**Fig. 17.** *The calculated pulse height distribution from a natural Si detector*.

Lastly, Figs. 16 and 17 show the pulse height distribution as a function of recoil energy, when a 100 kg natural Ge or natural Si detectors are placed 10 m away from the reactor. Overall, the pulse height distributions decrease when the nuclear recoil energies increase. The maximum threshold of recoil energies is different for both natural Ge and natural Si due to the difference in their masses.

## 5. Conclusion

In this work, we demonstrate the sensitivity of the reactor antineutrino spectrum in the calculation of the CE$\nu$NS detector response to a 1-MW research reactor at Texas A&M University. We show that the antineutrino spectrum from a nuclear reactor is very sensitive to the isomeric transitions of the various fission products isotopes. The reactor antineutrino spectrum is obtained using the summation method where the individual antineutrino spectra from all fission products undergoing beta-decay are summed. The detector responses and pulse height distributions are obtained by convolving the calculated CE$\nu$NS macroscopic cross-sections and the antineutrino spectrum emitted from the reactor.

The antineutrino spectrum without the isomeric transitions results in overestimation of detector response by more than 37 %. First, most isomeric transitions were included using the beta-decay data available in ENDF/B-VIII.0. Some beta intensity data are not available in the ENDF/B-VIII.0, and assuming 100 % beta intensity due to the ground state only, results in a 3.1 % error in the results. Thus, contributions of missing isomeric transitions were included using the Gross Theory. In our study, we have used the continuous beta spectrum from JENDL-2015 to fill-in those isotopes that don’t have beta intensity information in the ENDF/B-VIII.0 library. The finite size, radiative, and weak magnetism corrections result in a 0.8 % increment in the detector response. To mitigate the issue of the pandemonium effect, not possible using the ENDF/B-VIII.0 library for some isotopes, we included the TAGS data from published literature including the newest TAGS results from J. Gombass *et al.* (2021). Our analysis shows there is an 1.6 % decrease in the detector response by including the TAGS data set. In conclusion, isomeric transitions correction should be given priority when modelling the reactor antineutrino spectra, as their contributions to the detector response are high compared to others. On the other hand, Ishimoto's approximation method provides a similar result to the summation method such that only 0.25% and 0.38 % deviation is observed for natural Ge and natural Si response, respectively. We note that the approximation method using Ishimoto's approach can be a faster way to model antineutrino spectrum for preliminary research study purposes.

In this paper, we also demonstrate that natural Ge detector has better detection efficiency with 44.25 events/day when compared to 7.99 events/day from natural Si detector. However, the natural Si detector (0.51 MeV with 20 eV nuclear recoil threshold) is more sensitive at lower energies compared to the natural Ge detector (0.82 MeV with 20 eV nuclear recoil threshold). We present the probability distribution of recoil energies for a given incident antineutrino energy, where the probability of nuclear recoil energy is a decreasing function. The pulse shape distribution for both natural Ge and natural Si detectors also show a similar behaviour where the number of counts is higher at lower nuclear recoil energies. The study also demonstrates different materials have different recoil thresholds. Overall, lower mass nuclides have higher lower nuclear recoil thresholds in terms of minimum detectable incident antineutrino energy.

## 7. Appendix

**Table 7.** The TAGS data sets that used in this study.

| References | Isotopes |
|---|---|
| R.C. Greenwood *et al.* (1997) | $^{89}$Rb $^{90}$Rb $^{90m}$Rb $^{93}$Rb $^{93}$Sr $^{94}$Sr $^{94}$Y $^{95}$Sr $^{95}$Y $^{138}$Cs $^{138m}$Cs $^{139}$Cs $^{140}$Cs $^{141}$Cs $^{141}$Ba $^{142}$Ba $^{142}$La $^{143}$Ba $^{143}$La $^{144}$Ba $^{144}$La $^{145}$Ba $^{145}$La $^{145}$Ce $^{146}$Ce $^{146}$Pr $^{147}$Ce $^{147}$Pr $^{148}$Ce $^{148}$Pr $^{148m}$Pr $^{149}$Pr $^{149}$Nd $^{151}$Pr $^{151}$Nd $^{152}$Nd $^{152}$Pm $^{153}$Nd $^{153}$Pm $^{154}$Nd $^{154}$Pm $^{155}$Nd $^{155}$Pm $^{156}$Pm $^{157}$Pm $^{157}$Sm $^{158}$Sm $^{158}$Eu |
| A. Algora *et al.* (2010) | $^{102}$Tc $^{104}$Tc $^{105}$Tc $^{106}$Tc $^{107}$Tc $^{105}$Mo $^{101}$Nb |
| A. A. Zakari-Issoufou *et al.* (2015) | $^{92}$Rb |
| S. Rice *et al.* (2017) | $^{86}$Br $^{91}$Rb |
| L. Le Meur PhD thesis (2018) | $^{99}$Y $^{142}$Cs $^{138}$I |
| V. Guadilla *et al.* (2019) | $^{100}$Nb $^{100m}$Nb $^{102}$Nb $^{102m}$Nb |
| J. Gombas *et al.* (2021) | $^{103}$Nb $^{104m}$Nb |

**Table 8.** The beta continuous spectrum from JENDL-2015 that integrated into the calculation in this study.

| Atomic number (Z) | Isotopes |
|---|---|
| 29 | $^{71}$Cu |
| 30 | $^{74}$Zn $^{79}$Zn |
| 31 | $^{77}$Ga $^{82}$Ga $^{83}$Ga |
| 32 | $^{84}$Ge $^{85}$Ge $^{86}$Ge $^{87}$Ge |
| 33 | $^{83}$As $^{85}$As $^{86}$As $^{87}$As $^{88}$As $^{89}$As $^{90}$As |
| 34 | $^{88}$Se $^{89}$Se $^{90}$Se $^{91}$Se $^{92}$Se |
| 35 | $^{92}$Br $^{93}$Br $^{94}$Br $^{95}$Br |
| 36 | $^{94}$Kr $^{95}$Kr $^{96}$Kr $^{97}$Kr $^{98}$Kr |
| 37 | $^{99}$Rb $^{100}$Rb |
| 38 | $^{102}$Sr $^{103}$Sr |
| 39 | $^{100m}$Y $^{101}$Y $^{102}$Y $^{102m}$Y $^{103}$Y $^{104}$Y $^{105}$Y |
| 40 | $^{103}$Zr $^{104}$Zr $^{105}$Zr $^{106}$Zr $^{107}$Zr $^{108}$Zr |
| 41 | $^{104}$Nb $^{105}$Nb $^{106}$Nb $^{107}$Nb $^{108}$Nb $^{109}$Nb $^{110}$Nb $^{112}$Nb |
| 42 | $^{106}$Mo $^{107}$Mo $^{108}$Mo $^{109}$Mo $^{111}$Mo $^{112}$Mo $^{113}$Mo |
| 43 | $^{109}$Tc $^{110}$Tc $^{111}$Tc $^{112}$Tc $^{113}$Tc $^{114}$Tc $^{115}$Tc $^{116}$Tc |
| 44 | $^{111}$Ru $^{112}$Ru $^{114}$Ru $^{115}$Ru $^{116}$Ru $^{117}$Ru $^{118}$Ru |
| 45 | $^{112}$Rh $^{114}$Rh $^{115}$Rh $^{116}$Rh $^{117}$Rh $^{118}$Rh $^{119}$Rh $^{120}$Rh $^{121}$Rh |
| 46 | $^{115}$Pd $^{115m}$Pd $^{117}$Pd $^{119}$Pd $^{120}$Pd $^{121}$Pd $^{122}$Pd $^{123}$Pd $^{124}$Pd |
| 47 | $^{118}$Ag $^{119m}$Ag $^{120}$Ag $^{122m}$Ag $^{123}$Ag $^{124}$Ag $^{125}$Ag $^{126}$Ag $^{127}$Ag $^{128}$Ag |
| 48 | $^{124}$Cd $^{126}$Cd $^{127}$Cd $^{128}$Cd $^{129}$Cd $^{131}$Cd $^{132}$Cd |
| 49 | $^{113}$In $^{134}$In |
| 50 | $^{131}$Sn $^{131m}$Sn $^{135}$Sn $^{136}$Sn $^{137}$Sn |
| 51 | $^{129m}$Sb $^{136}$Sb $^{137}$Sb $^{138}$Sb $^{139}$Sb |
| 52 | $^{137}$Te $^{138}$Te $^{139}$Te $^{140}$Te $^{141}$Te $^{142}$Te |
| 53 | $^{140}$I $^{141}$I $^{142}$I $^{143}$I $^{144}$I |
| 54 | $^{142}$Xe $^{143}$Xe $^{144}$Xe $^{145}$Xe $^{146}$Xe $^{147}$Xe |
| 55 | $^{144}$Cs $^{146}$Cs $^{148}$Cs $^{149}$Cs |
| 56 | $^{148}$Ba $^{149}$Ba $^{150}$Ba $^{151}$Ba $^{152}$Ba |
| 57 | $^{149}$La $^{150}$La $^{151}$La $^{152}$La $^{153}$La $^{154}$La |
| 58 | $^{149}$Ce $^{150}$Ce $^{151}$Ce $^{152}$Ce $^{153}$Ce $^{154}$Ce $^{155}$Ce $^{156}$Ce |
| 59 | $^{153}$Pr $^{15}$4Pr $^{155}$Pr $^{156}$Pr $^{157}$Pr |
| 60 | $^{156}$Nd $^{157}$Nd $^{158}$Nd $^{159}$Nd $^{160}$Nd |
| 61 | $^{152n}$Pm $^{158}$Pm $^{159}$Pm $^{160}$Pm $^{161}$Pm |
| 62 | $^{160}$Sm $^{161}$Sm $^{162}$Sm $^{163}$Sm |
| 63 | $^{160}$Eu $^{161}$Eu $^{162}$Eu $^{163}$Eu $^{164}$Eu $^{165}$Eu |
| 64 | $^{163}$Gd $^{164}$Gd $^{165}$Gd $^{166}$Gd $^{167}$Gd |
| 65 | $^{166}$Tb $^{167}$Tb $^{168}$Tb $^{169}$Tb |
| 67 | $^{171}$Ho |